\documentclass[12pt,preprint]{aastex}
\def\r23{$R_{23}$}

\def\kms{km~s$^{-1}$}
\def\cm2{cm$^{-2}$}
\def\hkpc{$~h_{70}^{-1}$ kpc}

\shorttitle{Galaxy Pairs in the Sloan Digital Sky Survey I}
\shortauthors{Ellison et al.}

\begin{document}

%% LaTeX will automatically break titles if they run longer than
%% one line. However, you may use \\ to force a line break if
%% you desire.

\title{Galaxy Pairs in the Sloan Digital Sky Survey I: Star Formation,
AGN Fraction, and the Luminosity/Mass-Metallicity Relation.}

%% Use \author, \affil, and the \and command to format
%% author and affiliation information.
%% Note that \email has replaced the old \authoremail command
%% from AASTeX v4.0. You can use \email to mark an email address
%% anywhere in the paper, not just in the front matter.
%% As in the title, you can use \\ to force line breaks.
\author{Sara L. Ellison\altaffilmark{1},
David R. Patton\altaffilmark{2,3},
Luc Simard\altaffilmark{4},
\& 
Alan W. McConnachie\altaffilmark{1}
}

\altaffiltext{1}{Dept. of Physics \& Astronomy, University of Victoria,  
3800 Finnerty Rd, Victoria, V8P 1A1, British Columbia, Canada,
sarae@uvic.ca, alan@uvic.ca}

\altaffiltext{2}{Department of Physics \& Astronomy, Trent University, 
1600 West Bank Drive, Peterborough, Ontario, K9J 7B8, Canada, dpatton@trentu.ca}

\altaffiltext{3}{Visiting Researcher, Dept. of Physics \& Astronomy, 
University of Victoria,  
3800 Finnerty Rd, Victoria, V8P 1A1, British Columbia, Canada}

\altaffiltext{4}{National Research Council of Canada,
Herzberg Institute of Astrophysics, 5071 West
Saanich Road, Victoria, British Columbia, V9E 2E7, Canada,
luc.simard@nrc-cnrc.gc.ca}

%% Notice that each of these authors has alternate affiliations, which
%% are identified by the \altaffilmark after each name.  Specify alternate
%% affiliation information with \altaffiltext, with one command per each
%% affiliation.

%% Mark off your abstract in the ``abstract'' environment. In the manuscript
%% style, abstract will output a Received/Accepted line after the
%% title and affiliation information. No date will appear since the author
%% does not have this information. The dates will be filled in by the
%% editorial office after submission.

\begin{abstract}
We present a sample of 1716 galaxies with companions within $\Delta$v$<$
500 \kms, $r_p <$ 80 \hkpc\ and stellar mass ratio $0.1 <$ M$_1$/M$_2 < 10$
from the Sloan Digital Sky Survey (SDSS)
Data Release 4 (DR4).  The galaxy pairs are selected from the Main Galaxy
Sample using stringent and well-understood criteria for  redshift,
spectral quality, available stellar masses and metallicities.  
In agreement with previous studies, we find an enhancement in the
star formation rate (SFR) of galaxy pairs at projected separations 
$<$ 30--40 \hkpc.  
In addition, we find that this enhancement is highest (and extends 
to the greatest separations) for galaxies of approximately equal mass, the
so-called `major' pairs. However, SFR enhancement can still be detected
for a sample of galaxy pairs whose masses are within a factor of 10
of each other.  Based on these results, we define a sample
of close pairs ($\Delta$v$<$ 500 \kms, $r_p <$30 \hkpc,
$0.1 <$ M$_1$/M$_2 < 10$) which we use to investigate interaction 
induced effects in the luminosity-metallicity (LZ) relation.
In agreement with the one previous study of the LZ relation
in paired galaxies, we find an offset to lower
metallicities (by $\sim$ 0.1 dex) for a given luminosity for galaxies in pairs
compared to the control sample. We also present the first mass-metallicity
(MZ) relation comparison between paired galaxies and the field, 
and again find an offset to lower metallicities (by $\sim$ 0.05 dex)
for a given mass. The smaller offset in the MZ relation indicates that
both higher luminosities and lower metallicities may contribute
to the shift of pairs relative to the control in the LZ relation.
We show that the offset in the LZ relation
depends on galaxy half light radius, $r_h$.
Galaxies with $r_h \lesssim$ 3 \hkpc\ 
and with a close companion show a 0.05-0.1 dex downwards 
offset in metallicity compared to control galaxies of the same
size.  Larger galaxies do not show this offset
and have LZ and MZ relations consistent with the control sample. 
We investigate the physical impetus behind this empirical
dependence on $r_h$ and consider the galaxy's dynamical time
and bulge fractions as possible causes.  We conclude that the
former is unlikely to be a fundamental driver of the offset in
the LZ relation for paired galaxies, but that bulge fraction
may play a role.  Finally, we study the AGN fraction in both the
pair and control sample and find that whilst selecting galaxies
in different cuts of color and asymmetry yields different AGN
fractions, the fraction for pairs and the control sample are
consistent for a given set of selection criteria.  
This indicates that if AGN are
ignited as a result of interactions, this activity begins
later than the close pairs stage (i.e. once the merger is complete).
\end{abstract}

%% Keywords should appear after the \end{abstract} command. The uncommented
%% example has been keyed in ApJ style. See the instructions to authors
%% for the journal to which you are submitting your paper to determine
%% what keyword punctuation is appropriate.

\keywords{galaxies:abundances--galaxies:ISM}

%% From the front matter, we move on to the body of the paper.
%% In the first two sections, notice the use of the natbib \citep
%% and \citet commands to identify citations.  The citations are
%% tied to the reference list via symbolic KEYs. The KEY corresponds
%% to the KEY in the \bibitem in the reference list below. We have
%% chosen the first three characters of the first author's name plus
%% the last two numeral of the year of publication as our KEY for
%% each reference.

\section{Introduction}

The evolutionary path followed by a galaxy is shaped by its merger
history, which in turn depends on its environment.  This dependence
is epitomized by the properties of galaxies in rich
environments such as clusters (e.g. Dressler 1980; Whitmore, Gilmore
\& Jones 1993; Balogh et al. 1998, 1999; Poggianti et al. 1999;
Pimbblet et al. 2002; Wake et al. 2005).
The effect of such high density living is generally to suppress star formation,
through mechanisms that can include cluster tidal fields, 
gas (ram pressure) stripping, and strangulation (e.g. Byrd \& Valtonen 1990;
Moore et al. 1999; Diaferio et al 2001). From this
point of view, one may expect to see the most extreme effects of
density-induced properties from environments that are rich
on scales of a few hundred kpc.  
However, it is now emerging that density on smaller scales 
can be the major impetus behind galaxy evolution (e.g. Lewis et al. 2002;
Gomez et al. 2003; Blanton \& Berlind 2007).  
Galaxies in compact groups, for example,
exhibit a clear tendency towards lower metallicities and older
stellar populations compared with isolated galaxies (e.g. Proctor et al.
2004; Mendes de Oliveira et al. 2005; de la Rosa et al. 2007).
On these smaller scales, galaxy mergers provide the most obvious mechanism 
for change.  Simulations predict that prior to halted star formation,
there should be a phase of increased activity (e.g. Di Matteo,
Springel \& Hernquist 2005) which precedes the final merger, 
particularly in gas-rich systems.
Observations of early-stage galaxy interactions will therefore
complement those of rich environments to provide a more
complete picture of the evolutionary process.  In this sense,
close pairs or morphologically disturbed galaxies may be the
pre-cursors to the `red-and-dead' galaxies seen in dense 
environments.

The seminal study of the effect of interactions on galaxy colors
is the work of Larson \& Tinsley (1978).  
They found that disturbed galaxies in the
Arp catalogue had a wider spread of colors, including more
blue galaxies, than the field galaxies in the Hubble atlas.
In the last 30 years, this distinction in color has been confirmed
numerous times in larger samples. In general, galaxies with close companions,
including those showing clear signs of morphological asymmetry,
tend to have bluer (integrated)
optical colors (e.g. Carlberg et al. 1994; Patton
et al. 1997, 2005).  These results are indicative of enhanced
star formation, a scenario supported by high equivalent widths of H$\alpha$
emission when spectra are available (e.g. Kennicutt et al. 1987;
Barton, Geller \& Kenyon et al. 2000; Lambas et al. 2003;
Alonso et al. 2004; Nikolic, Cullen \& Alexander 2004).  
In turn, the star formation
heats galactic dust which emits thermally in the IR, leading
to an IR-excess in galaxy pairs (Kennicutt
et al. 1987; Xu \& Sulentic 1991; Geller et al. 2006).
This large body of observational data paints a clear picture of
enhanced star formation activity associated with galaxy proximity
on scales of a few tens of kpc.

Clues to the finer details of enhanced star formation can
be gleaned from galaxy simulations.
In the models of Mihos \& Hernquist (1994, 1996), the interaction-induced 
star formation occurs specifically in the central
regions (inner 1--2 kpc) of the galaxy, as a result of gas inflows.  
Observational evidence to support this theoretical prediction includes
1) centrally peaked distributions of H$\alpha$ and continuum emission in
interacting galaxies (Bushouse 1987; Smith et al. 2007), 2) enhanced H$\alpha$
flux or suppressed metallicities determined from nuclear spectroscopy 
of interacting galaxies 
(e.g. Barton et al. 2000; Kewley, Geller \& Barton 2006) and 3) 
enhanced radio continuum emission in the central parts of
pairs of spiral galaxies, but not in their disks (Hummel 1981).
However, there are also claims for enhanced disk star formation
(e.g. Kennicutt et al. 1987).  At the same time,
models of interacting galaxies predict the nature of induced
star formation to depend sensitively on the mass distribution
in the galaxies.  For example, for interactions that are observed
early-on in the merging process, Mihos \& Hernquist (1996) found
that galaxies with shallower potentials (i.e. less bulge dominated)
more efficiently funnel gas to the center through the formation of
a bar.  Conversely, bulge dominated galaxies are minimally affected
by close interactions until the merger event is well advanced
(e.g. Cox et al. 2007).

The induced star formation activity associated with interactions
and mergers is expected to have an impact on the metallicity of
galaxy pairs.  There is a well established correlation between
luminosity and metallicity, which is a manifestation of a more
fundamental stellar mass-metallicity relation (e.g. Tremonti et al. 2004;
Salzer et al. 2005; Lee et al. 2006), which is likely
be be `disturbed' for interacting galaxies.  It is not clear
\textit{a priori} how these scaling relations between luminosity, mass
and metallicity might be affected by interactions.  The galaxy luminosity may 
significantly increase due to the additional star formation
experienced as a result of the merger.  The overall metallicity
of an interacting galaxy may first appear to decrease as metal-poor
gas flows into its inner regions.  However, we eventually expect
the metallicity to increase as the star formation proceeeds and
eventually returns its nucleosynthetic products into the interstellar
medium.  The end point metallicity will depend on a number of factors
such as the mass and metallicity of the inflowing gas, efficiency of
the starburst and the metal yield.  The first major observational
study of these effects was presented by Kewley et al. (2006a),
who found a shift towards lower metallicities by $\sim$ 0.2 dex
in galaxy pairs for a given luminosity, compared to a control sample.
Since their spectra included only the central 10\% of the galaxies'
light, Kewley et al. (2006a) interpreted this result as the signature
of metal-poor gas that had been funnelled into the center of the
galaxies.

As the merging process advances,  an expected consequence of the
gas funnelling might be the
ignition of an AGN.  Effectively all galaxies are thought to harbor
black holes at their centers, the masses of which correlate
on the mass of the galaxy's bulge component as measured through
stellar velocity dispersion (e.g. Marconi et al.
2004; Shankar et al. 2004 and Ferrarese \& Ford 2005 for a review).  
Infall of
gas onto the black hole via a galaxy interaction is a natural
way to engage nuclear activity.  Indeed, it has previously been noted
that low redshift Seyfert galaxies often occur in groups
(e.g. Stauffer 1982) and that a high fraction of galaxies close
to AGN appear to be interacting (see the review by Barnes \&
Hernquist 1992).  However, although Seyfert galaxies may show evidence
for recent nuclear star formation (e.g. Storchi-Bergmann et al. 2001),
there is so far no evidence that AGN activity is enhanced in denser
environments relative to the field, including in close pairs
(e.g. Schmitt 2001; Sorrentino, Radovich \& Rifatto 2003; Alonso
et al. 2007).  Instead, AGN activity is best signalled by
morphological disturbances (e.g. Barnes \& Hernquist
1992; Alonso et al. 2007).

Investigating the myriad effects of galaxy interactions clearly 
requires measurements of a suite of properties, including stellar mass,
star formation rates (SFRs), AGN 
contribution, metallicities, color and morphology as characterized
by measures such as bulge-to-total ratios and asymmetry.  Whilst
many of these properties have been previously studied (see above
references) no work to date has been able to combine all of
these parameters for a single, large sample.
In this regard the Sloan Digital Sky Survey (SDSS)
is an excellent resource with both photometric and quality spectroscopic
data available for over half a million galaxies in the Data Release
4 (DR4).  In this paper series (see also Patton
et al in preparation, henceforth Paper II, 
and other forthcoming papers) we have combined SDSS 
photometry with the results of
spectral synthesis modelling, which yield estimates for
properties such as the stellar mass, metallicity, star formation
rate and bulge and disk image decomposition in five filters (Simard,
in preparation) to yield morphological parameters.  Therefore,
this sample provides the first coherent dataset for which such
a wide suite of galaxy parameters can be investigated, and the
relationships between these properties studied in a systematic way.
Moreover, the statistical power of the SDSS allows us 
to be highly selective
in the way we form our sample.  Therefore, although our final
pairs sample is not the largest to date (c.f. 
Alonso et al. 2006; Paper II), our selection criteria
are amongst the most stringent.  This is particularly important
when using spectroscopic data to determine quantities such as
metallicity, where the combination of several emission lines
can become very sensitive to poor S/N (e.g. Kewley \& Ellison,
2008). In Paper II we investigate the photometric
properties of SDSS galaxies in close pairs.  
%Our main results include the finding that
%galaxies with companions with transverse separations $<$ 30 \hkpc\ 
%and $\Delta v < 500$ \kms\ tend
%to have bluer, more dominant bulges and lumpier morphologies
%than the field population.
In this paper, we combine the basic survey properties of
a sample of galaxy pairs with spectroscopic properties determined by e.g.
Kauffmann et al. (2003b), Brinchmann et al. (2004), Tremonti et al. (2004) and
Kewley \& Ellison (2008).  This allows us to investigate
the sensitivity of metallicity, AGN incidence, mass and star
formation rate on a galaxy's proximity to a companion.

The layout of this paper is as follows.  In \S\ref{sample_sec}
we describe the compilation of our galaxy pairs and control
samples.  In \S\ref{sfr_sec} we use the wide pairs sample defined
in \S\ref{sample_sec} to study the effect of pair proximity
and relative stellar masses on star formation rate.  Based on
these results, we define a sample of close pairs which are
most likely to exhibit interaction-induced effects.  In \S\ref{met_sec}
we investigate the luminosity- and mass-metallicity relations
and in \S\ref{agn_sec} the AGN fraction in galaxies with close companions.
Each of the three science sections (\S\ref{sfr_sec} -- \ref{agn_sec})
can be read largely independently, although we recommend that
all readers understand the sample selection laid out in \S\ref{sample_sec}.  
We summarize the full results of this paper in \S\ref{summary_sec}.

We adopt a concordance cosmology of $\Omega_{\Lambda} = 0.7$, $\Omega_M = 0.3$,
$H_0 = 70$ km/s/Mpc.

\section{Sample Selection}\label{sample_sec}

Our galaxy pairs sample is selected from the DR4 of the SDSS and
includes requirements based on both photometric and spectroscopic
selection.  The imaging portion of the DR4 covers 6670 deg$^2$ in
five bands and the spectroscopic catalog is magnitude limited for
extinction corrected Petrosian $r<17.77$.  
To construct our galaxy samples, we use
the DR4 catalog of 567,486 galaxies compiled by the Munich group 
\footnote{http://www.mpa-garching.mpg.de/SDSS/}.  Pipeline
processing which fits galaxy templates and
spectral synthesis models to the spectra
yields physical properties such as stellar masses and star formation rates
as well as measurements of line fluxes (e.g. Kauffmann et al. 2003b;
Brinchmann et al. 2004; 
Tremonti et al. 2004).  Although metallicities are available
for the majority of these galaxies, Ellison \& Kewley (2005)
and Kewley \& Ellison (2008) have shown that
different empirical calibrations can yield metallicities that vary
by up to a factor of 3.  The Tremonti et al. (2004) metallicities
are amongst the highest of these calibrations.  We used the published
line fluxes to calculate the metallicities according to
the `recommended' method of Kewley \& Dopita (2002)
which solves iteratively for metallicity and ionization parameter.
We made this selection for two reasons.  First, the calibration of
Kewley \& Dopita (2002) yields one of the tightest mass-metallicity 
relations (Kewley \& Ellison 2008).  Second, the metallicity conversions
between various strong line diagnostics presented by Kewley \& Ellison
(2008) show that conversions to/from the Kewley \& Dopita (2002) 
calibration exhibit one of the smallest scatters.
Other properties used in this paper (e.g. SFR and
stellar mass) are taken directly from the catalogs made
generously available by the Munich team.  

Our sample selection differs
importantly from that of Paper II, which focuses on the photometric
properties of galaxies in pairs.  Although spectroscopic redshifts 
and stellar masses were
required for pair selection in Paper II, no other spectral requirement
was included in the selection criteria.  However, since we will
be focussed on properties that are derived from spectra, such as
SFR and metallicity, our selection criteria are more stringent,
and our sample correspondingly smaller.  Moreover, since our metallicity
determinations require moderately high S/N in the emission lines (see 
below), the galaxies in this sample are necessarily star-forming
or AGN dominated.  There are no quiescent, inactive (`red and dead')
galaxies in our sample.

From the catalog of over half a million SDSS DR4 galaxies we select 
galaxies that fulfill the following criteria:

\begin{enumerate}

\item Galaxies must have extinction corrected Petrosian
magnitudes in the range 14.5 $< r$ $\le$ 17.77.  The faint limit
matches the criterion of Sloan's Main Galaxy Sample and ensures
a high completeness and unbiased selection for mass estimates (see
below). The bright limit avoids deblending problems that confuse the
identification of close pairs (Strauss et al. 2002).
We also required that the objects were classified as galaxies from
the SDSS imaging (\textit{SpecPhoto.Type=3}) and were classified
spectrally as either a galaxy or QSO (\textit{SpecPhoto.SpecClass=2,3}).

\item Galaxies must be unique spectroscopic objects.  We reject duplicates
in the initial sample of 567,486 galaxies by including the single
galaxy that has been classified as `science worthy' (flag scienceprimary=1 
in the SDSS `SpecObjAll' table).  In a handful ($<20$) of cases, a single
photometric object (galaxy) is associated with two spectroscopic
objects.  Such cases are often highly disturbed galaxies which,
for example, have double nuclei or distinctive tidal tails.  We
reject these objects from our sample, but will re-visit them
in a future paper.

\item  The redshift must be $z < 0.16$ and the SDSS SpecObjAll parameter
which measures the redshift confidence $zconf > 0.7$.  
We exclude higher redshifts
since there is a tail of rare galaxies at $z>0.16$ which is not
seen in the pairs sample, simply due to small number statistics.
Imposing a redshift cut is common practice in pairs' studies in order
to limit the effects of both evolution and aperture effects
(e.g. Kewley et al. 2006; Woods \& Geller 2007), 
although we re-visit the redshift distribution at a more
sophisticated level below.

\item The error on the emission line flux must be less than one fifth
of the measured flux in all of the following emission lines: [OII]
$\lambda$3727, H$\beta$, [OIII] $\lambda$5007, H$\alpha$, [NII]
$\lambda$6584 and [SII] $\lambda \lambda$ 6717, 6731.  This criterion
ensures a high effective S/N which in turn facilitates accurate
classification of the galaxies as either star-forming or AGN-dominated
(e.g. Kewley et al.  2001; Kauffmann et al. 2003a) and for accurate
metallicity determination from empirical strong line diagnostics
(e.g., Kobulnicky, Kennicutt \& Pizagno 1999; Kewley \& Ellison, 2008).
This criterion automatically selects star-forming galaxies and will
exclude passively evolving or `red and dead' galaxies, as well as
galaxies with very high extinction and metal-poor galaxies with
faint emission lines.    

\item Stellar mass estimates must be available (e.g. Kauffmann
et al 2003b; Tremonti
et al. 2004).  These are available in the Munich catalogs and
are derived from spectral template fitting and have typical uncertainties
$\sim$ 0.1 dex.  Drory, Bender \& Hopp (2004)
have shown that the spectrally determined stellar 
masses compare well with those derived
from optical and IR colors and they are good surrogates for
the dynamical mass when log M$_{\star} > 10$M$_{\odot}$
(see also Brinchmann \& Ellis 2000).  At lower stellar masses,
M$_{\star}$ is larger than the dynamical mass by $<$ 0.4 dex
(Drory et al. 2004).

\item Metallicities as calculated by the Kewley \& Dopita (2002) 
diagnostic must be available, although we do not require that 
\textit{both} galaxies in a pair have known metallicities.  

\item Galaxies must be classified as star-forming and not AGN
dominated, according to the line diagnostic criteria given
in Kewley et al. (2001).  We impose this criterion since metallicities
derived from strong line calibrations assume a stellar ionizing
background and are not applicable if there is a (local) AGN component.
Recently, Kewley et al. (2006b) 
have proposed a new AGN removal scheme that is more stringent than
the original Kewley et al. (2001) criteria.  However,
Kewley \& Ellison (2008) have shown
that, for metallicities derived from the Kewley \& Dopita (2002)
strong line calibration, the mass metallicity relation is
identical for the Kewley et al. (2001) and Kewley et al. (2006b)
AGN filtering schemes.  We remove the criterion of AGN exclusion for
our study of AGN fractions in \S\ref{agn_sec}.\\

\noindent  From this master sample, we then select galaxies with
companions that we shall refer to as `galaxy pairs', although $\sim$ 5\%
consists of galaxies in triples and a minority of higher multiples.
For inclusion in the sample of galaxy pairs, we further require that

\item   Galaxies have one or more companions with projected
physical separations of $r_p < 80$ \hkpc.  
Although previous observational and theoretical
studies have found 30 \hkpc\ ($\sim$ 20 
$~h_{100}^{-1}$ kpc) to be the approximate scale on which
pairs start to exhibit distinct properties compared with the field 
(e.g. Barton et al. 2000; Patton et al 2000;
Lambas et al. 2003; Alonso et al. 2004;
Nikolic, Cullen \& Alexander 2004; Perez et al. 2006a), we consider
wider pairs in order to investigate trends in separation.
All pairs with separations $r_p < 15$ \hkpc\ were inspected visually,
since erroneous pair identifications do occur at small
separations.  The majority of spurious pairs were at a $r_p < 5$ \hkpc\
and occur e.g. when an HII region in a single galaxy is identified as 
separate galaxy.  For separations $r_p > 10$ \hkpc, the fraction
of spurious pair identifications is less than 1\%.

\item The rest-frame
velocity difference of a galaxy pair must be $\Delta v < 500$
\kms.  This velocity offset was selected in order to provide a
balance between contamination and statistics.  Although a much smaller
velocity separation reduces contamination, it also reduces the
overall sample size, which may ultimately become a limiting 
factor in pair statistics. The trade-off between these effects
has been addressed in Patton et al. (2000).  

\item Relative stellar masses must be within a factor of 10.  
Although we expect to see more interaction-induced
effects in pairs of almost equal mass
(e.g. Woods, Geller \& Barton 2006; Cox et al. 2007; Woods
\& Geller 2007), we
include a wide range of mass ratios in order to investigate
the relative impact of major and minor interactions.

\end{enumerate}

\begin{figure*}
\centerline{\rotatebox{0}{\resizebox{13cm}{!}
{\includegraphics{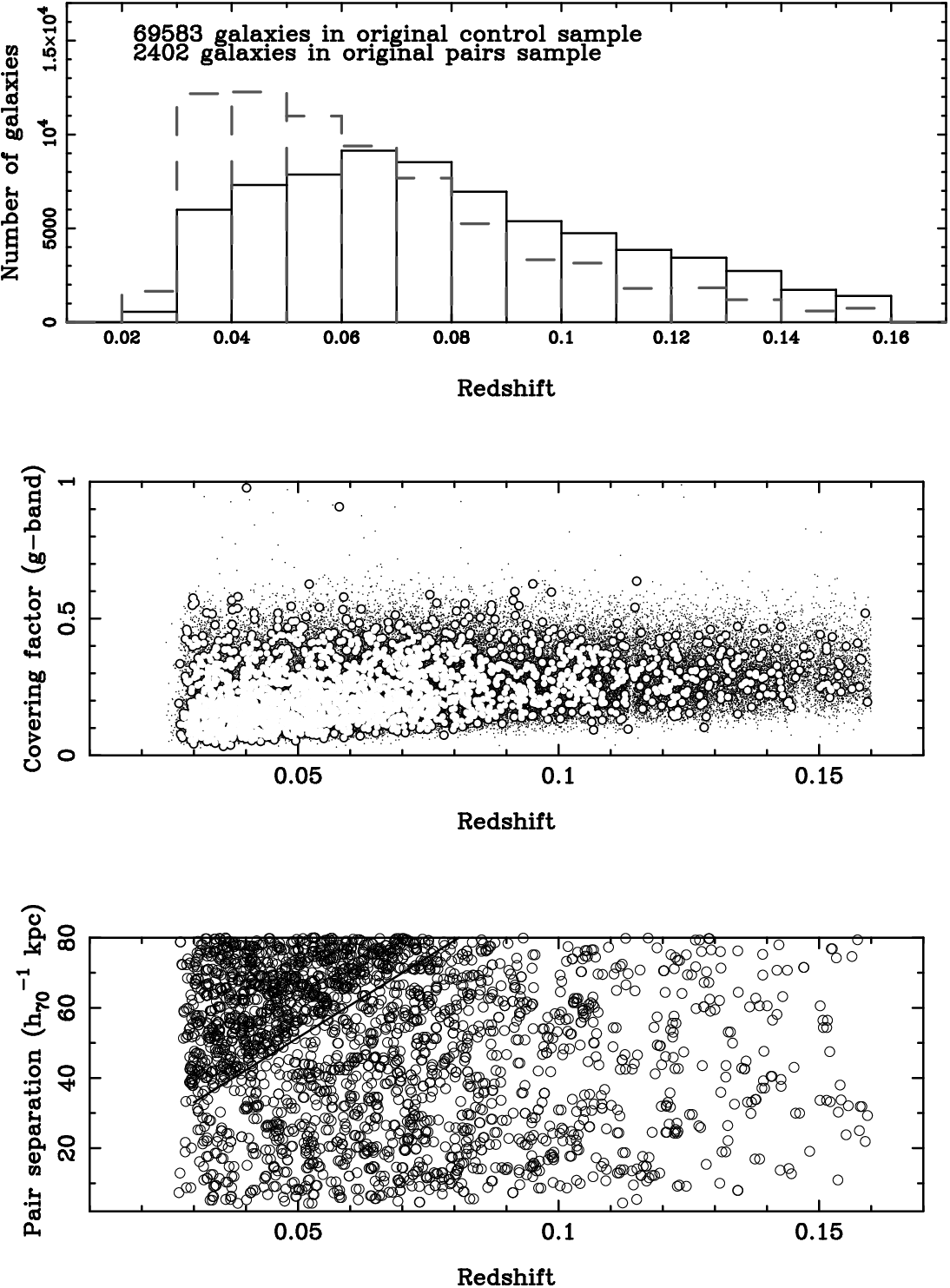}}}}
\caption{\label{z_histo_nocull} Top panel:  Redshift histogram 
of the \textit{candidate} control galaxy 
sample (solid) and \textit{candidate} wide pairs sample (dashed),
with the latter scaled for display purposes.  
Middle panel: Covering fraction as a function of redshift for our
control (black dots) and pairs (open circles) samples. Bottom panel: 
Separations of galaxy pairs as a function of redshift (symbols as
before).  
The diagonal solid line shows a constant 
angular separation of $55\arcsec$, corresponding to the SDSS
fiber collision limit.}
\end{figure*}

If a galaxy fulfills the first seven of the above criteria, but not
the latter three, it is a candidate for our control sample.  
A galaxy fulfilling all ten criteria may be potentially included 
in our sample of wide pairs.  Before constructing the final control
and wide pairs samples, we make two further restrictions in order
to make the two samples directly comparable.  Both of these restrictions
are driven by the requirement that the redshift and stellar 
mass distributions of
the pairs and control samples should be statistically indistinguishable.
This is an important requirement since the distributions of
stellar mass and redshift can impact the observed ranges in
properties such as luminosity and star formation
rate.  The redshifts of the galaxies selected simply from the above criteria
are shown in Figure \ref{z_histo_nocull} where the histogram of 
pairs' redshifts has been roughly normalized to the number of 
galaxies in the control sample for display purposes.
Clearly, the redshift distribution
of the pairs is skewed towards lower values than the control,
which could potentially bias our results.  This can largely be understood
by examining the lower panel of Figure \ref{z_histo_nocull} which
shows the projected 
separation of pairs as a function of redshift and demonstrates a
clear excess of pairs at low redshift and wide separation.  This
is mostly due to the spectroscopic follow-up strategy of the SDSS survey.
There is a 55 arcsecond fiber collision limit due to the size of the
fiber housing, which prevents pairs with angular separations less 
than this from
being observed spectroscopically on the same plate\footnote{At $z=0.05$,
1 arcsec $\sim$ 1 \hkpc}.  However,
contiguous plates have considerable overlap and some sky regions 
are observed more than once, so that many close
pairs exist in the final spectroscopic catalog.  
The net effect on our preliminary pairs sample is that the spectroscopic 
completeness drops sharply below $55\arcsec$, leading to the relative 
over-abundance of pairs with wide physical separations at low redshifts 
in Figure  \ref{z_histo_nocull}.
Fortunately, it is straightforward to model and correct for this effect.  
Patton \& Atfield (2008) find that
the ratio of spectroscopic to photometric pairs decreases from $\sim$ 80\% 
at angular separations $\theta > 55\arcsec$ to $\sim$ 26\% (on average) at 
smaller separations.  
We therefore make a first attempt to correct the disparity in redshift 
distributions by randomly excluding $54/80$ = 67.5\% of galaxies
in pairs with $\theta > 55\arcsec$.
We use this cull to compile our final wide pairs sample,
which contains 1915 paired galaxies before AGN removal 
and 1716 galaxies with one or more companions after AGN removal.

\begin{figure*}
\centerline{\rotatebox{0}{\resizebox{13cm}{!}
{\includegraphics{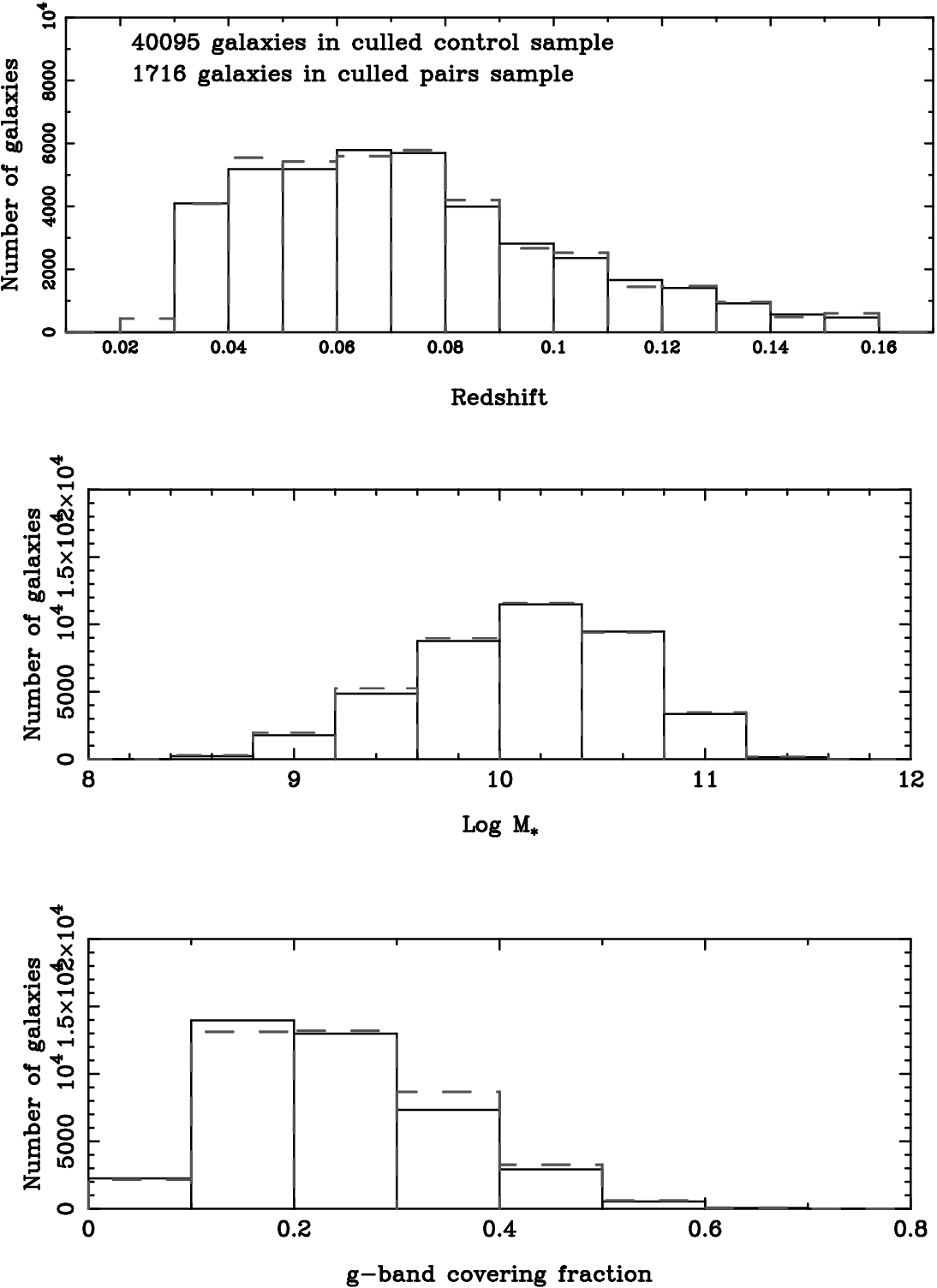}}}}
\caption{\label{z_histo_cull}
Histograms of redshift (top panel), mass (middle panel) and $g$-band
covering fraction (bottom panel) of our final control galaxy (solid) and wide 
pairs (dashed) samples, with the latter scaled for display purposes.
  }
\end{figure*}

When the stellar mass ratio of the pairs is not highly discrepant
($0.3< $M$_1$/M$_2 \lesssim 3$) the cull described above yields
redshift distributions for the pairs and control samples that are
statistically indistinguishable.  However, for more contrasting mass
ratios, the redshift distributions remain statistically different.
This is a common, well-known feature of pairs' samples (e.g., Patton
et al. 2000; 2005) and is due to the magnitude limited nature of the
parent galaxy sample and the associated limit in dynamic range.  Pairs
with very disparate stellar mass ratios are biased towards low
redshifts, because the magnitude limit of the survey hinders their
detection (i.e. detection of a much lower mass, fainter companion) at
higher redshifts.  Since we want to be able to study pairs with
stellar mass ratios up to 10, the control sample requires further
culling.  At this point, a simple prune in redshift is insufficient,
due to the strong correlation between mass and redshift.  At $z
\lesssim 0.05$ galaxies with stellar masses ranging from approximately
10$^{8.5}$ to 10$^{11}$ M$_{\odot}$ are detected.  
At higher redshifts, the lower
mass galaxies are no longer detected, since they are generally too
faint.  We therefore have to prune the control sample simultaneously
in stellar mass and redshift.  This is achieved by matching one
control galaxy to each paired galaxy in mass-redshift space and
repeating (without replacement) as many times as possible while
requiring that the KS probability of the
control--pair mass and redshift distributions be consistent with each
other at at least the 30\% level.  The matching process is done before
any removal of AGN-dominated galaxies 
so that the analysis of \S\ref{agn_sec} (on AGN
fractions) can be achieved.  For each of the 1915 (pre-AGN removal)
paired galaxies,
there are 23 control galaxies, i.e. the control sample contains 44045
galaxies before AGN-dominated galaxies are removed.  
The KS probabilities that these samples of pairs and
control galaxies are indistinguishable in redshift and stellar mass is
32\% and 34\% respectively (i.e. no formal statistical difference).
Once the AGN-dominated galaxies have been removed, as is required for
the majority of our analysis, the samples are reduced to 1716 paired
galaxies and 40095 control galaxies, a reduction in each case by
approximately 10\%.  Figure \ref{z_histo_cull} shows the redshift and
stellar mass distributions for these fiducial samples.  For both
paired and control samples, the mean stellar mass is log $M_{\star}$ 
= 10.1 and the mean
redshift is $z=0.073$.  We can now be confident that our control
sample is well-matched to our pairs sample and should contain no
observational bias that will affect our assessment of proximity
induced effects.

The strict selection criteria that we impose mean that our sample
of galaxies is not complete in either magnitude or volume.
As noted above, the S/N criterion in particular will lead to
a sample that excludes (at least some) galaxies that are highly
reddened, very metal-poor and not actively star-forming.   However, the
same selection biases will apply equally to the control and
the pairs samples, allowing us to make differential comparisons
between the two.  
As described in the above discussion, our sample of pairs is 
also not complete.
This should not introduce any bias into our pairs sample, since 
spectroscopic incompleteness does not depend significantly on the 
intrinsic properties of the galaxies.  
However, due to spectroscopic incompleteness, many true 
pairs will have a redshift measured for only one member galaxy, which 
may then fall into the control sample.  Fortunately, any resulting 
contamination of the control sample is negligible, since only $\sim 2$\% 
of galaxies are found in close pairs (see Patton \& Atfield 2008).

Given that we are interested in the effects of mergers/interactions, 
it is also important to acknowledge the fact that some of the pairs in our 
sample will not be close enough for such encounters to occur.
This contamination is on the order of 50\% 
for the closer pairs  ($r_p < 30$ \hkpc) in our sample 
(Patton \& Atfield 2008), and rises as pair separation increases 
(Alonso et al. 2004, Perez et al. 2006a).  While we do not attempt 
to correct for this explicitly, we infer that 
(a) any differences seen between close pairs and the control sample 
are likely to be underestimated and (b) the wider pairs are likely 
to suffer from increasing contamination due to non-interacting systems.

One other parameter that may affect measured spectral properties
is the fiber covering fraction (CF).
Although aperture effects are likely to affect galaxy metallicities
(Kewley, Jansen \& Geller 2005; Ellison \& Kewley 2005) we 
do not make any \textit{a priori} cuts in CF.  This is
mainly because, after the above culls, the CFs are consistent
between the pairs and control samples, see Figure \ref{z_histo_cull}.
Moreover, we will be explicitly investigating 
the impact of CF,  which is calculated 
by comparing the galaxy's photometric g' band Petrosian magnitude
with the fiber magnitude in the same filter, as a free parameter
in sections \ref{sfr_sec} and \ref{met_sec}.  However, we note that 
the quantities of stellar mass and SFR are corrected for aperture
effects (Brinchmann et al. 2004) and therefore represent total
quantities.

\begin{figure*}
\centerline{\rotatebox{270}{\resizebox{13cm}{!}
{\includegraphics{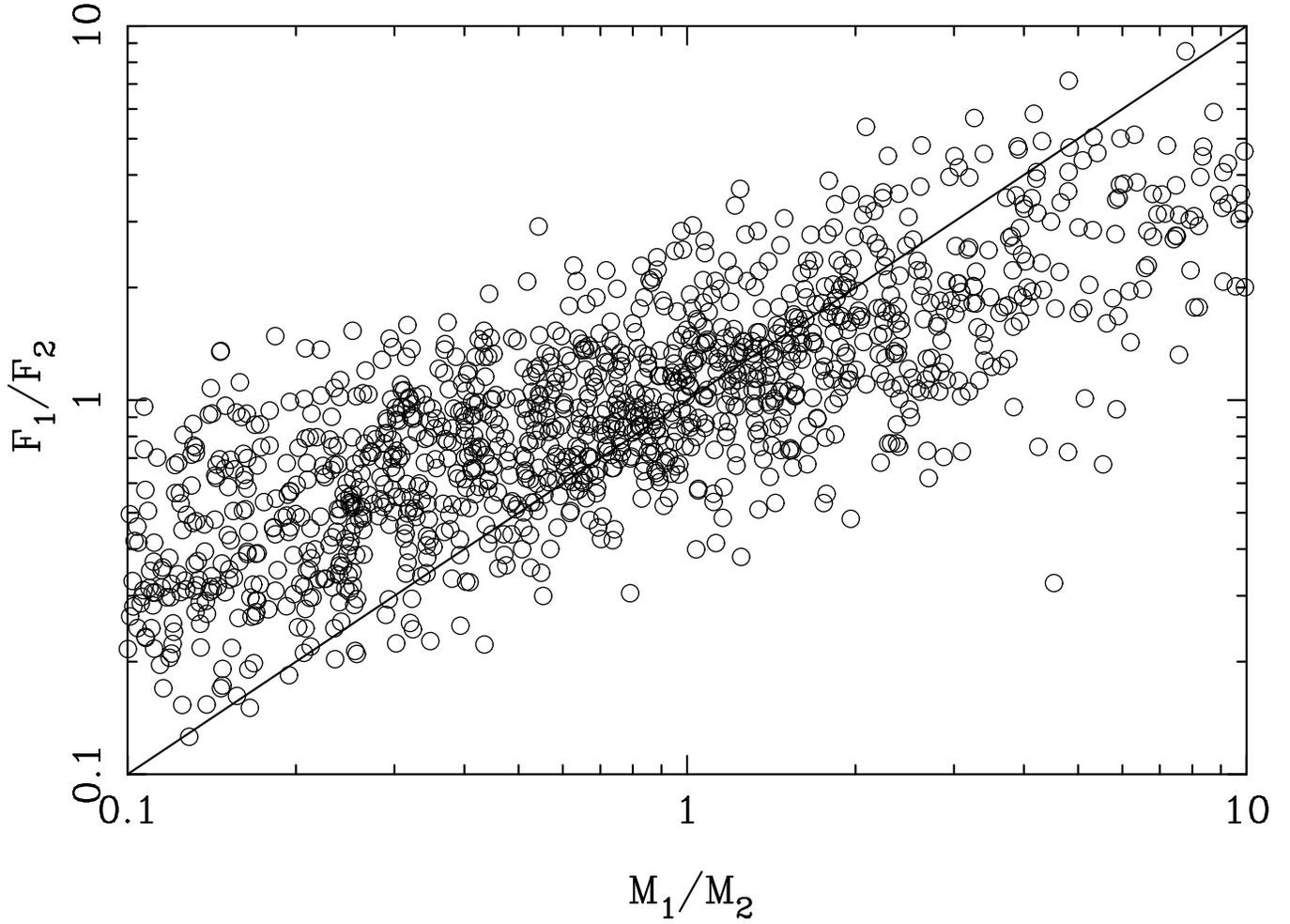}}}}
\caption{\label{m1m2_f1f2}Comparison between relative r-band
fluxes and relative masses in the wide pairs sample. The solid
diagonal line shows a one-to-one relationship between flux and mass ratios.}
\end{figure*}

It is worth noting that one of the novel properties of our sample
is the stellar mass selection criterion, whereas most previous samples 
have made no requirement on the relative fluxes/masses of their
visually identified pairs.  Moreover, when cuts have been made
in order to investigate the effect of relative mass, flux is
usually used as a surrogate for mass (e.g. Woods et al.
2006; Woods \& Geller 2007). 
In Figure \ref{m1m2_f1f2} we show the impact
of this assumption by plotting relative fluxes versus relative stellar
masses.  The fact that the distribution of flux to mass ratios
is flatter than 1:1, means that for a given flux ratio cut
the completeness rate for the same mass ratio is quite high,
but the contamination is significant.  For example, a flux selection
which requires a ratio within 2:1 selects 86\% of galaxy pairs with
masses whose ratios are within 2:1.  However, 46\% of the galaxies selected
by this flux cut will have actual mass ratios outside the 2:1 range, leading to
a high contamination rate.  Selection by relative flux could therefore
potentially dilute properties that depend sensitively on relative stellar mass.
The reason that the correlation between relative fluxes and masses
is flatter than unity in  Figure \ref{m1m2_f1f2} can be understood in terms of
specific star formation rates (SSFR).  Recently, Zheng et al. (2007)
have shown convincingly that SSFR, i.e. SFR per unit mass, is higher
for lower mass galaxies.  In turn, this broadly translates to a higher flux
per stellar mass (F/M) for lower mass galaxies.  Therefore, when the
M$_1$/M$_2$ ratio is less than unity, i.e. the low mass galaxy is
in the numerator, this translates to a generally
higher F$_1$/F$_2$, because F$_1$/M$_1 > $F$_2$/M$_2$.

With the stringent criteria outlined above, we have not only
constructed one of the largest, but also one of the
most rigorously selected samples of 
galaxy pairs to date.  Moreover, with the combination of a wide range
of derived spectral properties, photometric measurements and
morphological decomposition, we have an extensive arsenal with
which to tackle the effects of galaxy proximity.

\section{Star Formation Rate In Galaxy Pairs}\label{sfr_sec}

In this section, we investigate the effects of projected separation,
relative stellar masses and fiber covering fraction on the SFR of
paired galaxies.  We use the results to select a pairs sample
for the investigation of proximity effects on the LZ and MZ relations
in the following section.

In the top panel of Figure \ref{sfr_m1m2} we show the star formation rate
as a function of galaxy separation for the wide pairs in our sample.    
The figure demonstrates that galaxies in pairs with separations
$\le$ 30 \hkpc\ have a median SFR that is
higher than the control galaxies, by up to 40\%, at 1--2 $\sigma$ 
significance.   This result is
consistent with previous studies of SFRs in close pairs of
galaxies (e.g. Barton et al. 2000; Lambas et al. 2003;
Nikolic et al. 2004; Geller et al. 2006).   However, Barton
et al. (2008) have suggested that the level of excess star formation
in close pairs may have been under-estimated in these previous
works due to the typically higher density environments inhabited
by pairs relative to control galaxies.  This conclusion may
also apply to this work, although we discuss this further in the
next subsection. 

\subsection{Star Formation and Relative Galactic Stellar Mass}

We expect (e.g. Lambas et al. 2003; Woods et al. 2006; Bekki, Shioya
\& Whiting 2006; Cox et al. 2007; Woods \& Geller 2007) that pairs
with almost equal masses (`major mergers/interactions') will exhibit
more pronounced interaction-induced effects than unequal (`minor')
mass encounters.  Although dynamical mass may be the fundamental
parameter which governs the outcome of galaxy interactions, stellar
mass is both more readily determined from observations and is a
reasonable surrogate for dynamical mass above $10^{10}$ M$_{\odot}$
(e.g. Brinchmann \& Ellis
2000; Drory et al. 2004).  Moreover, stellar mass is a quantity that
is directly traced through many simulations, e.g. the minor pairs
models of Cox et al. (2007).

Only a handful of simulations have studied the effect of star formation
in minor mergers either in general (Mihos \& Hernquist 1994; Cox et al., 
2007) or for specific cases (e.g. Mastropietro et al. 2005
for the Milky Way -- LMC).  Based on this limited modelling, it has
been found that induced central star formation in the larger galaxy 
of an unequal mass merger can \textit{eventually} occur, albeit at a 
lower level than expected
in a major merger, and usually when the interaction is well advanced, i.e. 
after several gigayears.    In this section, we investigate whether
unequal mass pairs can be affected by galaxy proximity and compare 
our results with pairs whose galaxies have comparable stellar masses.
The only previous observational studies to assess the effects in
minor mergers in close galaxy pairs were those of
Woods et al. (2006) and Woods \& Geller (2007).  
The latter paper, which benefits from significantly better statistics
than the former, finds that the specific SFR of the less massive (as inferred
from a fainter magnitude) galaxy in a minor pair is enhanced compared
to the field, whereas the more massive galaxy is not.
However, these two previous studies relied upon
relative magnitudes, and as we pointed out in \S\ref{sample_sec},
this can lead to a high rate of contamination.
In this work, we use the measured stellar masses, corrected for
aperture bias, determined by spectral modelling and compare our results
to the flux-selected minor pairs of Woods \& Geller (2007)

\begin{figure*}
\centerline{\rotatebox{0}{\resizebox{13cm}{!}
{\includegraphics{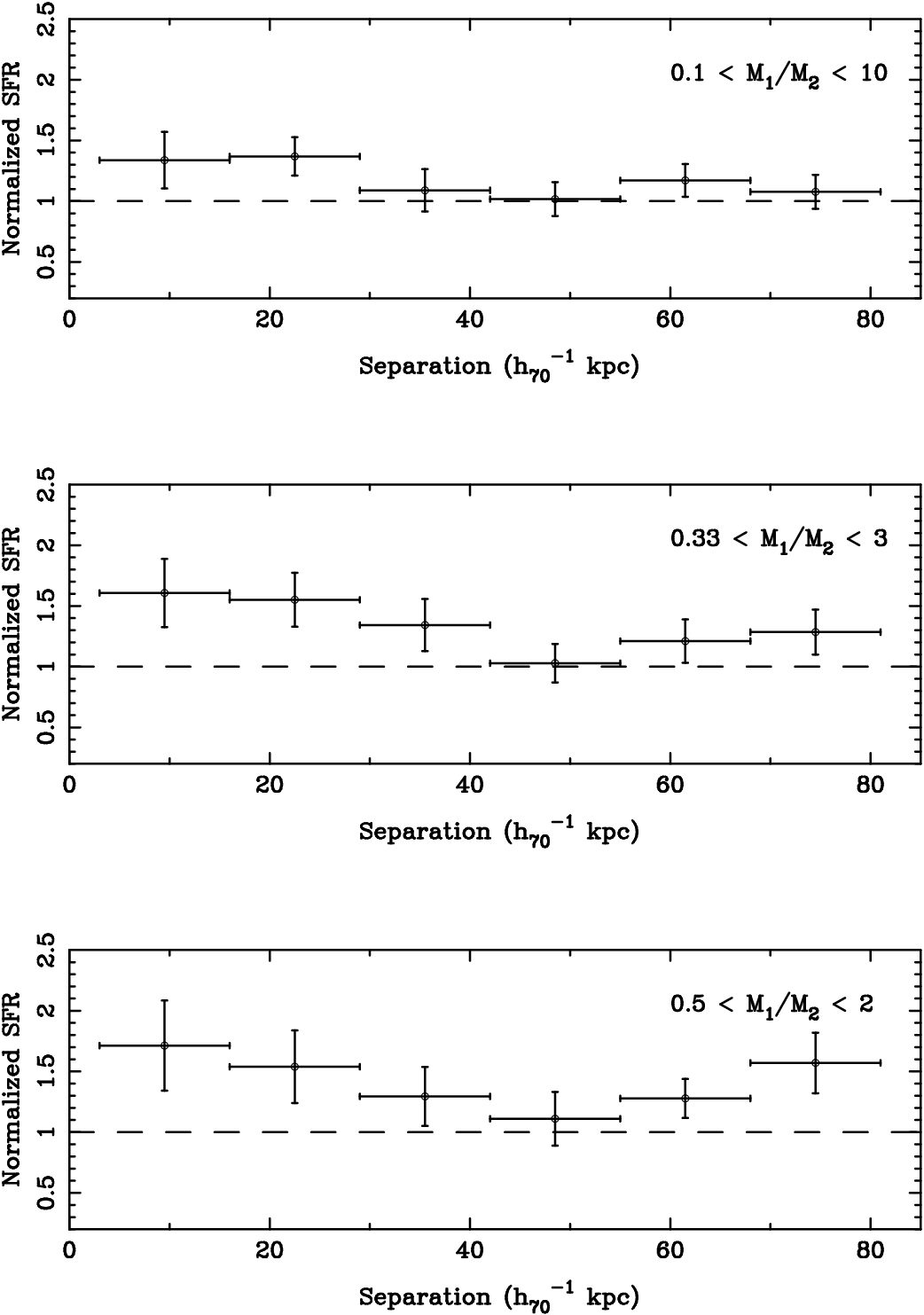}}}}
\caption{\label{sfr_m1m2} SFR for galaxies with a companion, as
a function of pair separation for three different mass ratio samples.
The SFRs have all been normalized to the median control value for
that mass range.  This figure
shows an increase in SFR relative to the field for projected separations
$r_p < 30$ \hkpc\ for all mass ratios.  As the disparity in masses
decreases (from top panel to bottom), this enhancement increases in
magnitude, significance and out to larger separations.  The apparent
increase in SFR at $r_p > 50$ \hkpc\ is due to contamination effects,
see text for details.}
\end{figure*}

We begin by assessing the impact of our mass
ratio criterion of $0.1 <$ M$_1$:M$_2 <10$ by considering
sub-samples of galaxy pairs with different stellar mass ratios.
For each mass cut, the matching of the control sample in
stellar mass and redshift is repeated as described in section \ref{sample_sec}
for our fiducial (wide) pairs sample.  This ensures that the distribution
of stellar masses is comparable between each pairs' sub-sample and its
control sample.
In Figure \ref{sfr_m1m2} we show the SFR as a function of
separation for three different mass ranges (stellar mass ratios within
1:10, 1:3 and 1:2).  From this Figure we draw two conclusions.
First, the enhancement in SFR persists out to at least 30 \hkpc\ for all
three mass ranges considered, with the closer stellar mass ratio pairs showing
an increase out to 40 \hkpc.  Second, and perhaps more interesting, is
that the amount of SFR enhancement increases (and becomes more
significant) for pairs whose stellar masses are most similar to one another.

The enhancement, which we found in the previous subsection to be 40\%
for pairs with stellar mass ratios within 1:10, increases to 60\% and
70\% for ratios within 3:1 and 2:1 respectively and with $\sim 2
\sigma$ significance in each case.  This confirms
quantitatively the suggestion that major interactions, i.e. those
between almost equal mass galaxies, will induce the most significant
effects in one another.  These results also demonstrate that the SFR
can be affected even in samples with relatively discrepant masses, at
least up to a ratio of 1:10 (as also concluded by Woods \& Geller 2007
for their minor pairs).  At large separations ($r_p \ga 50$ \hkpc) we
see an upturn in the SFR of the pairs.  This is a complex effect that
is driven by a combination of contamination from projected pairs that
are not truly interacting and the way in which our control sample is
constructed.  Since the control sample has been culled in redshift and
mass in order to match the distribution in the pairs sample, it is not
representative of the true field population.  Since pairs tend to be
found in higher density environments (e.g. Barton et al. 2008),
the mass-matched control sample has a higher mean
stellar mass than the field (i.e. the pre-cull control sample).  In
turn, this means that the control sample galaxies are themselves
biased towards denser environments and are therefore likely to have,
on average, lower SFRs than the field.  At wide separations, an
increasing number of pairs are not truly interacting, leading to an
increased contamination of the sample.  The SFRs at these wide
separations are therefore averages of the values of true interacting
pairs (which at wide separations probably have SFRs tending towards
the control mean) and contaminating field galaxies (which tend to have
a higher SFRs than the control mean).  This leads to an apparent
upturn in the SFRs at wide separations.  The prominence of this upturn
will depend on the actual mass matching of each mass ratio sub-sample.
Since the 0.1 $<$ M$_1$/M$_2 < $ 10 mass-matched control sample is
most similar to the field sample (i.e. mass distribution of the
pre-cull control sample), the upturn is much smaller (in fact, absent)
than in the 0.5 $<$ M$_1$/M$_2 < $ 2 mass-matched control sample,
which is most discrepant from the field mass distribution.  This
explains why the majority of previous surveys have not seen this upturn:
they do not impose relative flux or mass cuts, hence their
SFR versus separation correlations most closely approximate to the
top panel of Figure \ref{sfr_m1m2}.  For example, Lambas
et al. (2003) see an upturn at $r_p > 60$ kpc for their
$L_1 \sim L_2$ sample, but not in their $L_1 >> L_2$ sample.
Nonetheless, an upturn such as
that seen in the middle and bottom panels of Figure \ref{sfr_m1m2}
has been reported by Perez et al. (2006a) in their
analysis of mock galaxy pair catalogs from cosmological simulations,
by Nikolic et al. (2004) in their study of SDSS pairs and, as mentioned
above, by Lambas et al. (2003).  
We conclude that in the absence of projection effects, the SFR of
pairs with $r_p > 50$ \hkpc\ would tend to the control value.

Next, we classify major pairs as those with mass ratios within 2:1 and
minor pairs as those with more discrepant masses\footnote{This is
different to the definition of Woods \& Geller
(2007) who considered the major/minor boundary as a 2 magnitude difference
(factor 6.25) in brightness.}.  We further distinguish
between the more massive galaxy in a minor pair 
($M_{\rm gal} / M_{\rm companion}
> 2$) and the less massive galaxy in a minor pair 
($M_{\rm gal} / M_{\rm companion}
< 0.5$).  Since a number of fundamental galaxy
properties such as SFR depend on stellar mass (e.g. Brinchmann et al. 2004), 
simply using the matched control sample for comparison with minor/major 
pairs (whose mass distributions will be very different from one
another) would not give a true indication of relative effects.
We have therefore further adapted our control samples to be equivalent
in mass distribution by selecting a control galaxy matched in
stellar mass to each paired galaxy.

\begin{figure*}
\centerline{\rotatebox{270}{\resizebox{13cm}{!}
{\includegraphics{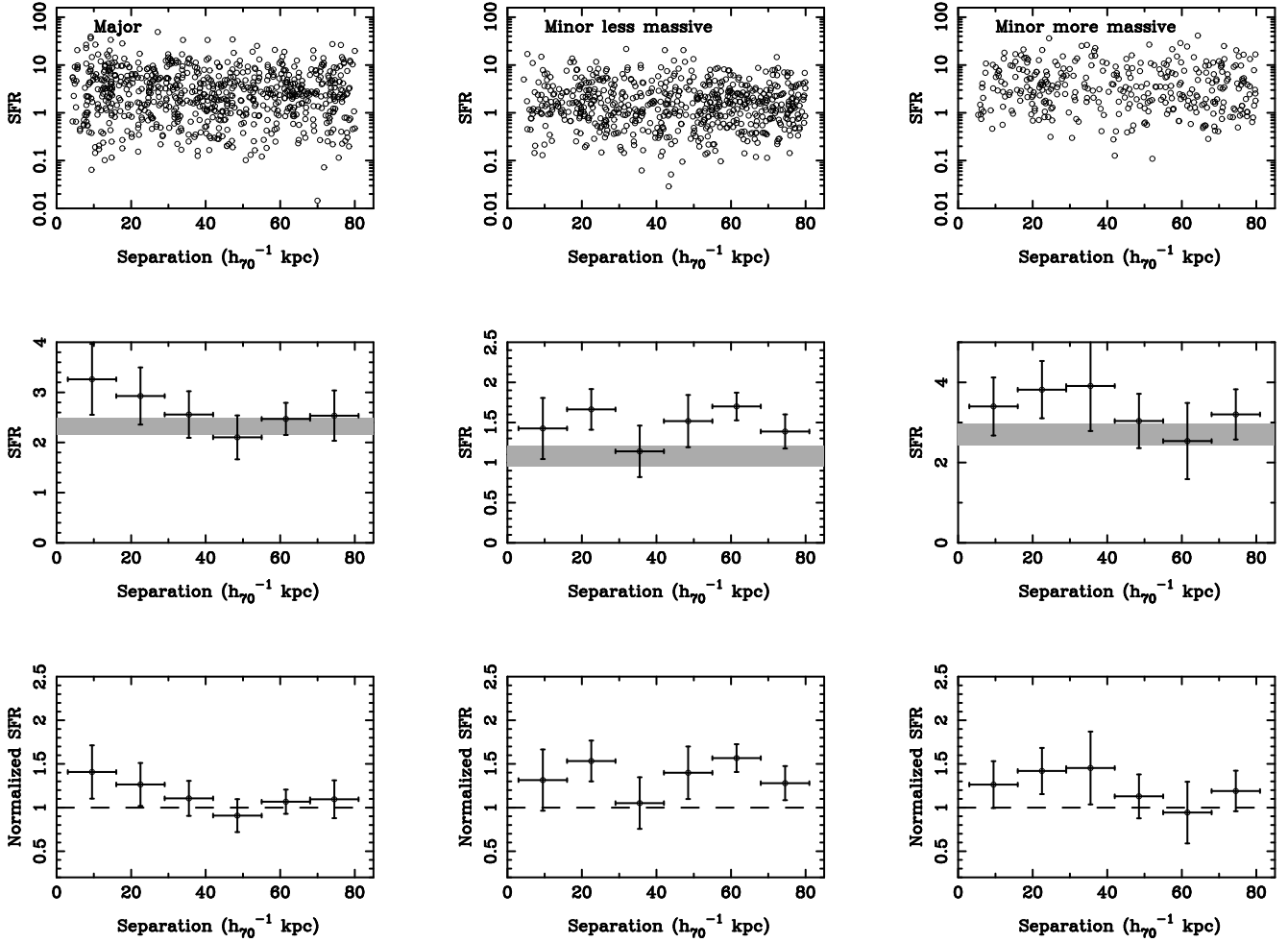}}}}
\caption{\label{sfr_mass} SFR for galaxies with a companion, as
a function of pair separation for three different mass samples.
The first column of 3 panels represents major pairs (mass ratio $<$ 2:1, 
the middle 3
are the less massive galaxies in minor pairs and the right-most 
column are the more massive galaxies in minor pairs.
Upper panels: individual galaxies. middle panels: SFRs binned by separation.  
The gray region shows the field median for galaxies
with a matched mass distribution (see text for details). 
Lower panel: SFRs normalized to the median control value.
The left-hand panels for major mergers confirm the results of Figure
\ref{sfr_m1m2}.  We find no convincing evidence for
enhanced SFR in minor mass pairs.}
\end{figure*}

In Figure \ref{sfr_mass} we show the total SFR as a function
of galaxy pair separation for 3 stellar mass scenarios:   major pairs
and the more/less massive galaxies in minor pairs.  
In the top panels of we show individual
SFRs, and in the middle panel, the median values in bins of
13 \hkpc.   The shaded region in the middle panel
shows the median 
SFR (with vertical height corresponding to the $\sigma/\sqrt{N}$)
in the matched control sample.
The overlap of the scatter in the
data points (vertical error bars on the binned values) with the gray
bar gives an indication of consistency with field values.
In the lower panel we show the SFR enhancement relative to the
control sample by normalizing each bin to the control median. 
The stellar mass matching of the control
fields is particularly important here.
It can be seen that the median values for the three middle
panels are highest for the highest mass sub-samples\footnote{
Whilst the \textit{specific} SFR is higher for lower mass
galaxies (e.g. Brinchmann et al. 2004) 
the total SFR is higher for higher mass galaxies}.  

Figure \ref{sfr_mass} demonstrates the, by now, familiar enhancement
of SFR at small separations for galaxies with approximately equal
masses, see also Figure \ref{sfr_m1m2}.  Although the result is not
highly significant ($\sim 1-2 \sigma$), we also find tentative
evidence for higher SFR for the less massive galaxy in a pair at both
close separations and at $\sim$ 60--70 \hkpc\ (see the previous
section for discussion on the turn-around in enhanced SFR as a
function of separation).  A similar conclusion has been drawn by Woods
\& Geller (2007).  Although some of the binned SDSS data points for
the more massive galaxy in a pair are also above the field mean, the
size of the error bars makes this result less significant (barely $1
\sigma$) and difficult to draw conclusions from.  If confirmed, these
results would be consistent with the less massive pair member in an
unequal mass interaction being susceptible to enhanced star formation,
although less so than galaxies in equal mass interactions.  In turn,
this result has interesting implications for cosmic metal enrichment:
Whereas low mass galaxies can usually remain gas-rich because of low
star formation efficiency, strong bursts of star formation during
interaction may increase metal production which may be more easily
dispersed into the surrounding intergalactic medium. However, the
results from this section are inconclusive and the analysis of Woods
\& Geller (2007) remain the strongest evidence for enhanced star
formation in less massive galaxies in minor pairs.  In a complementary
study of star-forming galaxies in the SDSS, Li et al.
(2008a) have also recently found evidence that SFRs are more
enhanced in lower mass galaxies with companions. Possible reasons
that we have not found similarly significant results include 1) the
different definition of major and minor pairs and 2) the smaller
sample size of our work, mostly due to the criteria imposed in \S
\ref{sample_sec} (although Woods \& Geller 2007 use the somewhat
larger DR5, compared to our DR4 sample).  The major pairs sample of
Woods \& Geller (2007) is 60\% larger than ours, whilst the minor
pairs sample is contains almost twice the number of galaxies.  The
median luminosity ratio for the Woods \& Geller minor sample is $\sim$
11 (compared with a median mass ratio of 3.85 in our sample), and
$\sim$ 4 for the major sample (compared with our median major mass
ratio of 1.38).  Therefore, if luminosity ratio were taken as a
substitute for mass ratio, more than half of the Woods \& Geller
(2007) major pairs sample would fall into our definition of a minor
pair.  In future work, it will be interesting to examine how selection
based on relative stellar masses and luminosities (e.g. Figure
\ref{m1m2_f1f2}) and the definition of major and minor pairs may affect
results.

Our data confirm the conclusion of previous
work (e.g. Barton et al. 2000; Lambas et al. 2003; Alonso
et al. 2004; Nikolic, Cullen \& Alexander 2004; Li et al. 2008a) that 
galaxies in pairs closer than $\sim$ 30 \hkpc\ exhibit SFRs
that are higher than in the `field'.  For the rest of
this paper, we therefore define a sample of `close pairs' where
$r_p < 30$ \hkpc.  Although we have shown that approximately equal
stellar mass pairs show higher proximity-induced SFRs, we elect to
use the $0.1<$ M$_1$/M$_2 < 10$ sample in order to maximise the
statistical significance of our work.  This selection also facilitates
comparisons with previous works, which generally do not have relative
stellar mass or flux limits in their pairs selection.
The $r_p <30$ \hkpc, $\Delta v <$ 500 \kms\ and  $0.1<$ M$_1$/M$_2 < 10$
criteria now forms our fiducial pairs sample unless otherwise stated.
The merging timescale for these galaxies  
is $\sim$ 250 -- 500 Myrs (e.g. Patton et al. 2000; Masjedi et al. 2006).

\section{Metallicities of Galaxy Pairs}\label{met_sec}

In the previous section, we used SFR as a function of separation
to define a `close pairs' sample with $r_h < 30$ \hkpc\ as those
pairs most likely to exhibit interaction induced effects.
We now use this sample to investigate the impact of proximity
on galaxy metallicity using this close pairs sample.

The metallicities of the SDSS galaxies can be determined using
strong emission line diagnostics that are calibrated either
empirically against `direct' electron temperature determinations,
or against theoretical photoionization models.  A wide range
of such metallicity diagnostics is currently on the market,
some of the most popular include various empirical calibrations of 
$R_{23}$ originally formulated by Pagel et al. (1979) (e.g. McGaugh 1991;
Zaritsky, Kennicutt \& Huchra 1994) empirical [NII]/H$\alpha$
calibrations (e.g. Denicolo, Terlevich \& Terlevich 2002;
Pettini \& Pagel 2004) and calibrations which solve iteratively
for ionization parameter using photoionzation models (e.g.
Kewley \& Dopita 2002; Kobulnicky \& Kewley 2004).  It is well
known that at high metallicities these strong line diagnostics
show a positive offset relative to the metallicities determined from electron
temperature methods (e.g. Bresolin, Garnett \& Kennicutt 2004;
Bresolin 2007).  Moreover, Kewley \& Ellison (2008) have shown that
strong, systematic differences exist between strong line
diagnostics and have stressed the importance of using a single
calibration where possible.  In this paper, we use the Kewley
\& Dopita (2002) `recommended' method which can both overcome the
usual double-value degeneracy of the R$_{23}$ method, and also
solves for the ionization parameter.

As noted in \S \ref{sample_sec} the (necessary) selection of galaxies
with strong emission lines means that our sample contains a dearth
of metal-poor galaxies.  However, not only does the consistent
selection of control and paired galaxies ensure an internally fair comparison,
but repeating our analysis with less stringent emission line detection
constraints (3$\sigma$ rather than 5$\sigma$) yields identical
results for all of the tests performed in this section.

\subsection{The Luminosity Metallicity Relation}\label{LZ_sec}

The relationship between luminosity and metallicity is well-established
over 8 magnitudes in M$_B$ (e.g. Salzer et al. 2005; Lee et al. 2006) 
and out to redshifts $z \sim$ 1 (e.g. Kobulnicky \& Kewley 2004; Maier
et al. 2005). 
The reason for the luminosity-metallicity (LZ) relation,
and the tighter mass-metallicity (MZ) relation is still unclear.
Although yields from luminous, high mass galaxies indicate that
the relation is driven by the depth of the potential well and
mass loss during star formation
(e.g. Tremonti et al. 2004), lower mass galaxies show
a large scatter in effective yield, with some showing values
as high as the most massive galaxies (Lee et al. 2006).
Simulations of chemical evolution offer a variety of alternatives,
including variable initial mass functions (Koppen, Weidner \& Kroupa
2007), star formation efficiency (Brooks et al. 2007) and the
interplay between poor-metal gas inflow and mass-loaded winds
(Finlator \& Dav\'e 2007).  Ellison et al. (2008) have recently
shown that the normalization of the MZ relation depends on
specific SFR and $r_h$, and conclude that differences in star formation
efficiencies can explain these dependencies.  
However, the basic form of the relation
remains intact over the full range in these properties and apparently
does not depend sensitively
on large-scale environment (Mouhcine, Baldry \& Bamford 2007).  

Regardless of the origin of the LZ relation, the enhanced star formation
discussed in the previous section should ultimately impact on the
correlation of luminosity and metallicity in close galaxy pairs.
The direction of this impact is dependent on timescales.  If a galaxy's
metallicity is measured after an interaction-driven starburst is
complete, then we may expect an enhanced metallicity in the HII regions
where star formation has occurred.  Conversely, if we
measure the metallicity of the region experiencing the starburst
whilst it is ongoing, the inflow of more metal-poor gas from the
outer regions of the galaxy may decrease the HII region metallicity.
Shifts in luminosity may also be applicable due to enhanced star formation.
This question has been recently tackled by Kewley et al. (2006a)
who used 86 galaxies in pairs selected from the CfA2 redshift survey
and compared them with a control sample from the Nearby Field Galaxy
Survey (NFGS).  For both samples, nuclear spectra containing $\sim$ 10\% of 
the galaxy's light were used for the metallicity determinations.
Kewley et al. (2006a) found that galaxies with separations $<$ 30 \hkpc\
\footnote{We have converted the separations used by Kewley et al.,
who quote distances in units of $~h_{100}^{-1}$ kpc, to our
cosmology and convention of \hkpc} have metallicities that are
offset downwards by 0.2 dex at a given luminosity.  Further observational
evidence that merger-induced starbursts lead to lower metallicities
comes from studies of ultra-luminous infra-red galaxies (ULIRGs; Rupke,
Veilleux \& Baker 2007) and compact ultra-violet luminous galaxies
(UVLGs; Hoopes et al. 2008).  These populations,
believed to have been recently involved in merger events, 
are more metal-poor by up to a factor of two compared to SDSS galaxies
of the same mass.  The simulations
of Perez et al. (2006b) also support the concept of metal-poor
gas inflow in pairs.  They find that the gas phase metallicity of
galaxies in simulated pairs is typically 0.2 dex higher when the
integrated metallicity over 2 optical radii is compared to that
over half an optical radius.

\begin{figure*}
\centerline{\rotatebox{270}{\resizebox{13cm}{!}
{\includegraphics{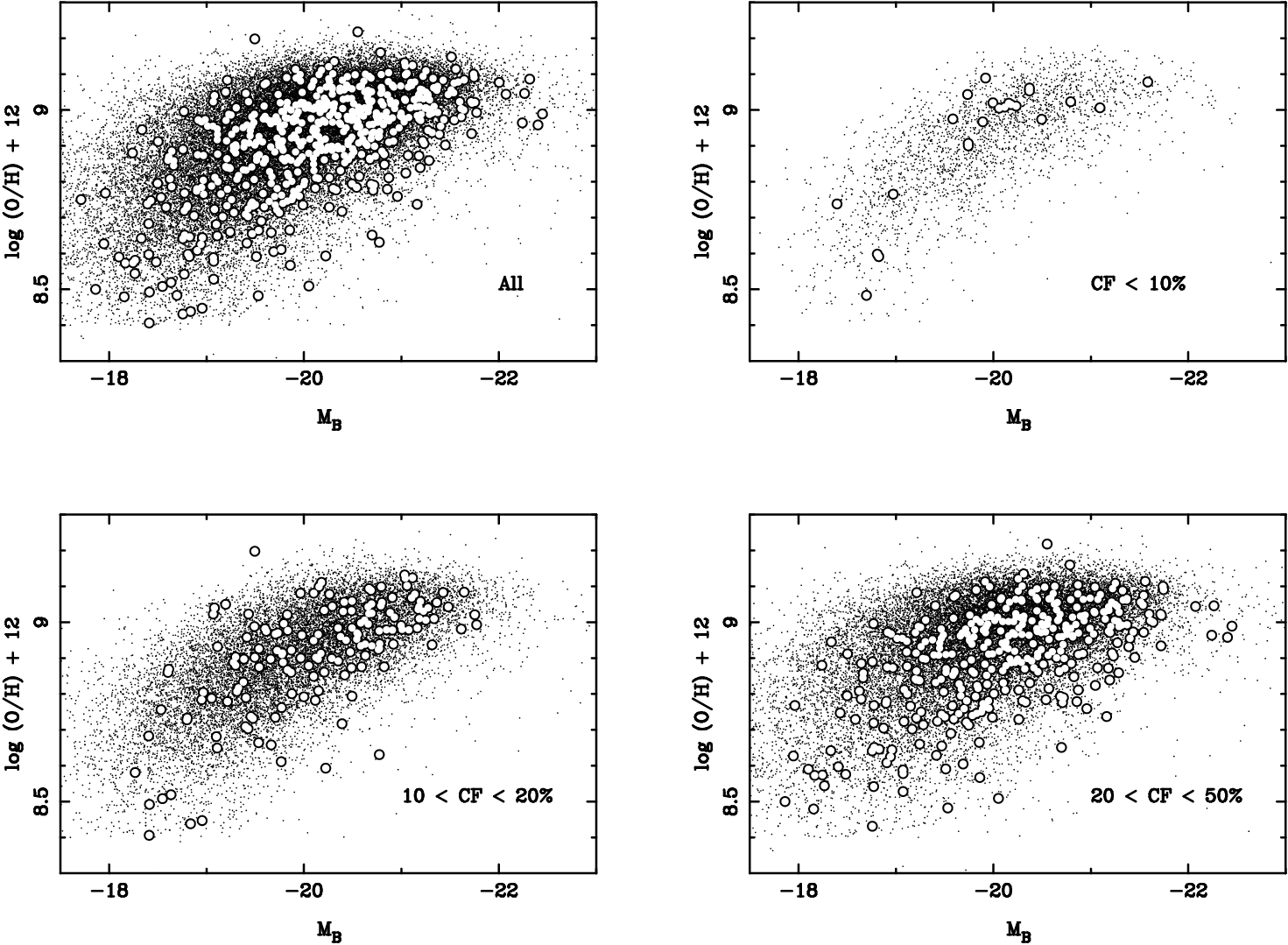}}}}
\caption{\label{LZ} Luminosity-metallicity (LZ) relations for our control
sample (black dots) compared with galaxies with close companions 
(open points) for various
cuts in $g$-band fiber covering fraction (CF).  }
\end{figure*}

\begin{figure*}
\centerline{\rotatebox{0}{\resizebox{13cm}{!}
{\includegraphics{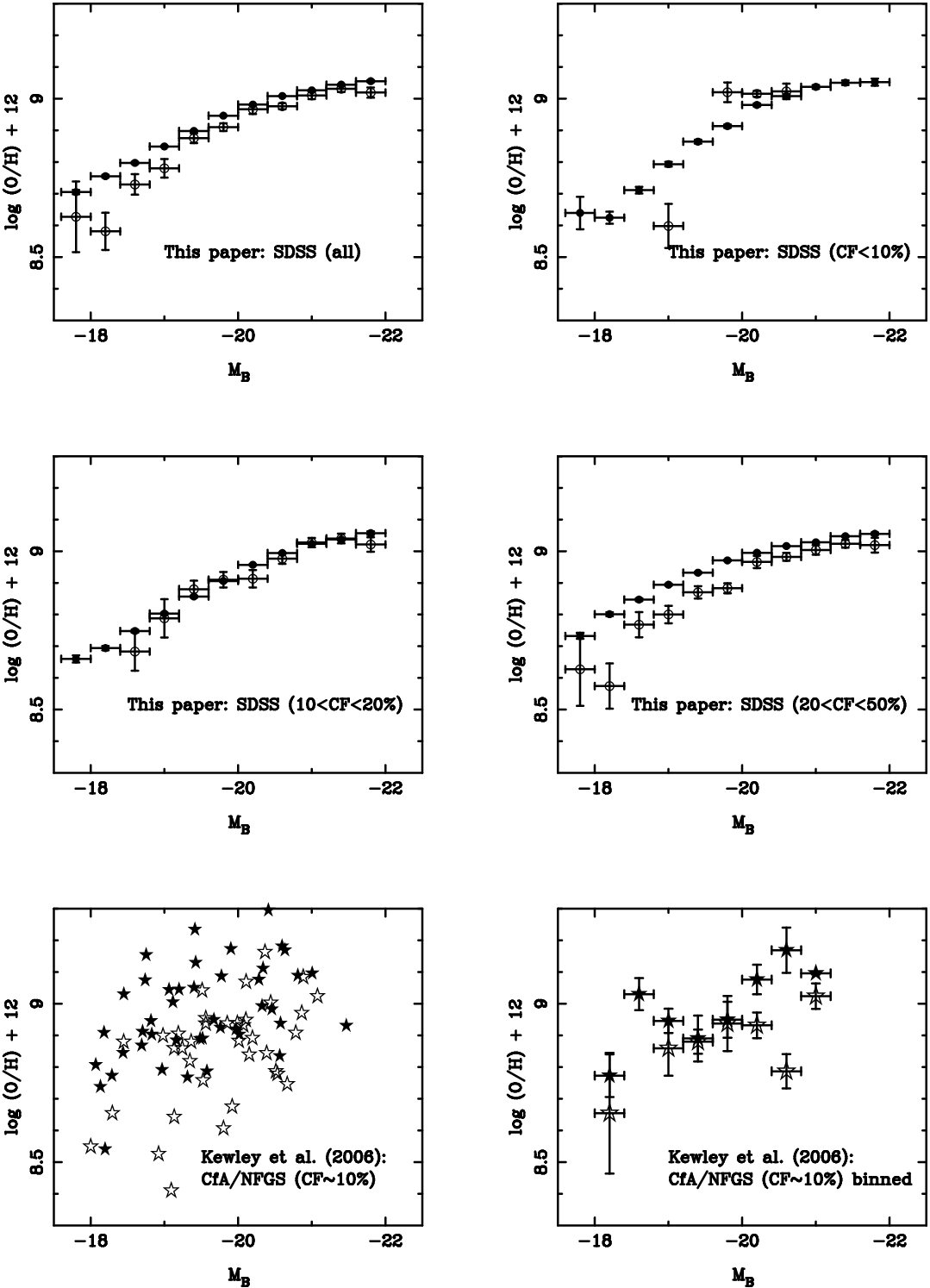}}}}
\caption{\label{LZ_bin} Luminosity-metallicity (LZ) relations for our
SDSS pairs sample (circles) and the CfA/NFGS sample (stars)
of Kewley et al. (2006). In all panels, filled points refer to 
control samples and open points to pairs. For the SDSS sample, we
show both the full control/pairs sample and CF cuts as in Figure \ref{LZ}
(see panel labels).  The Kewley
et al. (2006) data also correspond to CF$\sim$10\% and are shown
both unbinned (bottom left panel) and binned (bottom right panel). }
\end{figure*}

In Figure \ref{LZ} we show the LZ relation for our SDSS samples of
pairs and control galaxies.  In the top left panel we show all
galaxies in our close pairs sample, i.e. with transverse projected
separations $r_p < 30$ \hkpc.  Kewley et al. (2006a) have argued that
offsets from the field LZ relation will be most clear when the spectra
are of a nuclear nature, i.e.  only cover the central few kpc of the
galaxy where the starburst is occuring.  We therefore plot the LZ
relation for three different CF cuts.  In order to make any offsets
between the field and pairs more clear, in Figure \ref{LZ_bin}, we
show binned versions of all the SDSS pairs, as well as for the various
CF cuts. We note that due to the exclusion of very metal-poor galaxies
in our sample, it is possible that any downward shift in metallicity
in the pairs sample is under-estimated.   
For comparison, we also show the Kewley et
al. (2006a) CfA pairs sample and their NFGS control sample, both as
individual galaxies and binned.  The visual impression that the CfA
pairs of Kewley et al. (2006a) have lower metallicities for their
luminosity than NFGS control galaxies is confirmed quantitatively with
a 2D KS test which shows that the LZ distribution of the two samples
differs at the 98\% confidence level.

If we consider the SDSS sample as a whole (top left panels of
Figures \ref{LZ} and \ref{LZ_bin}), we see a mild tendency
towards lower metallicities for pairs compared with the control
sample.  However, the offset is small, $<$0.05 dex, compared to
the offset seen by Kewley et al. (2006a), which is typically
0.1--0.2 dex.  The main difference between
the SDSS sample and the NFGS/CfA sample studied by Kewley
et al. (2006a) is that the latter had
nuclear spectra with CF$ \sim 10$\%.  The majority
of the SDSS galaxies have much higher covering fractions (Figure
\ref{z_histo_cull}).  If the effect observed by Kewley et al.
(2006a) is therefore truly nuclear,
then the typically higher covering fractions of the SDSS fibers
may hide the impact of gas dilution in the galaxies' centers.
It would therefore be more appropriate to consider only the SDSS galaxies
(in both pairs and control sample) with CF$<$10\%.  The top right
panels in Figures \ref{LZ} and \ref{LZ_bin} show the individual galaxies,
and binned metallicities for the CF$<$10\% criterion.  Although
our sample of CF$<$10\% pairs is smaller than the CfA (23 galaxies, 
compared with 37 in the CfA), the scatter in metallicity is
also smaller for a given M$_B$, leading to smaller error bars
(which represent the standard error on the mean).  The SDSS 
CF$<$10\% control sample is also much larger than the NFGS: 2060
galaxies compared with 43 at comparable separations.  Figure \ref{LZ_bin}
therefore shows the interesting result that, at least for intermediate
luminosity galaxies, SDSS pairs with CF$<$10\% have marginally
higher metallicities
for their luminosity than the control sample.  Recall that this offset 
is in the opposite sense to the CfA pairs studied by Kewley et al. (2006a).  
A 2D KS test gives a 3\% probability that the SDSS control and pairs 
sample have the same LZ distributions.  As stated above, the KS probability 
is 2\% for the Kewley et al. samples, so both datasets give statistically
significant results, but in contrary directions.
It is worth noting that the covering fractions for the CfA and SDSS
samples are calculated slightly differently: Kewley et al. consider the
fraction of light in the slit relative to the $B26$ isophote, whereas
we consider the fiber magnitude relative to the Petrosian magnitude
in the $g$-band.  However, this can not explain the trend
of our result, i.e., that we see a larger offset in the pairs' LZ
relation relative to the control for higher CFs, which is contrary
to the expectation from nuclear metallicity dilution.

\subsection{Comparison with the work of Kewley et al. (2006a)}\label{ljk_sec}

The results in  the previous subsection indicate an
apparent discrepancy in the relative metallicities of galaxy pairs
in the SDSS versus the CfA samples for nuclear (CF$<$10\%)
spectra.  On the one hand,
Kewley et al. (2006a) find low metallicities at a given luminosity
in close pairs, whereas we find tentative evidence for
high metallicities compared
with a control sample when the CF$<$10\%.  Conversely, we do find
lower metallicities in pairs when the CF$>$20\% (Figure
\ref{LZ_bin}), a regime in which Kewley et al. (2006a)
have little data.  In
this subsection we investigate the cause of this apparent discrepancy.

First, we consider whether the small number of low CF galaxies
in the SDSS (23, versus 37 in the CfA) could lead to disagreement
relative to the nuclear LZ relation of Kewley et al. (2006a).    
We quantify the effect of small number statistics by 
bootstrapping 10,000 samples of 23 galaxy pairs
from the CfA sample and calculating the 2D KS probability compared
with the NFGS control sample.  This test simulates the effects of the
smaller number of pairs in the SDSS compared with the CfA, i.e.
by testing whether the CfA/NFGS comparison would have detected
an LZ offset if it had only had as many pair galaxies as the
small CF bin of the SDSS.  We find that for samples of 23 pairs
a significant KS probability of $<$0.05 is achieved in 86\% of the
bootstrap renditions 
and a probability of $<$0.02 for 63\% of trials.  Therefore,
although we can not completely rule out the possibility that
small numbers are the cause of the apparent discrepancy between
the SDSS and CfA nuclear LZ relation for pairs, it seems unlikely.

\begin{figure*}
\centerline{\rotatebox{0}{\resizebox{13cm}{!}
{\includegraphics{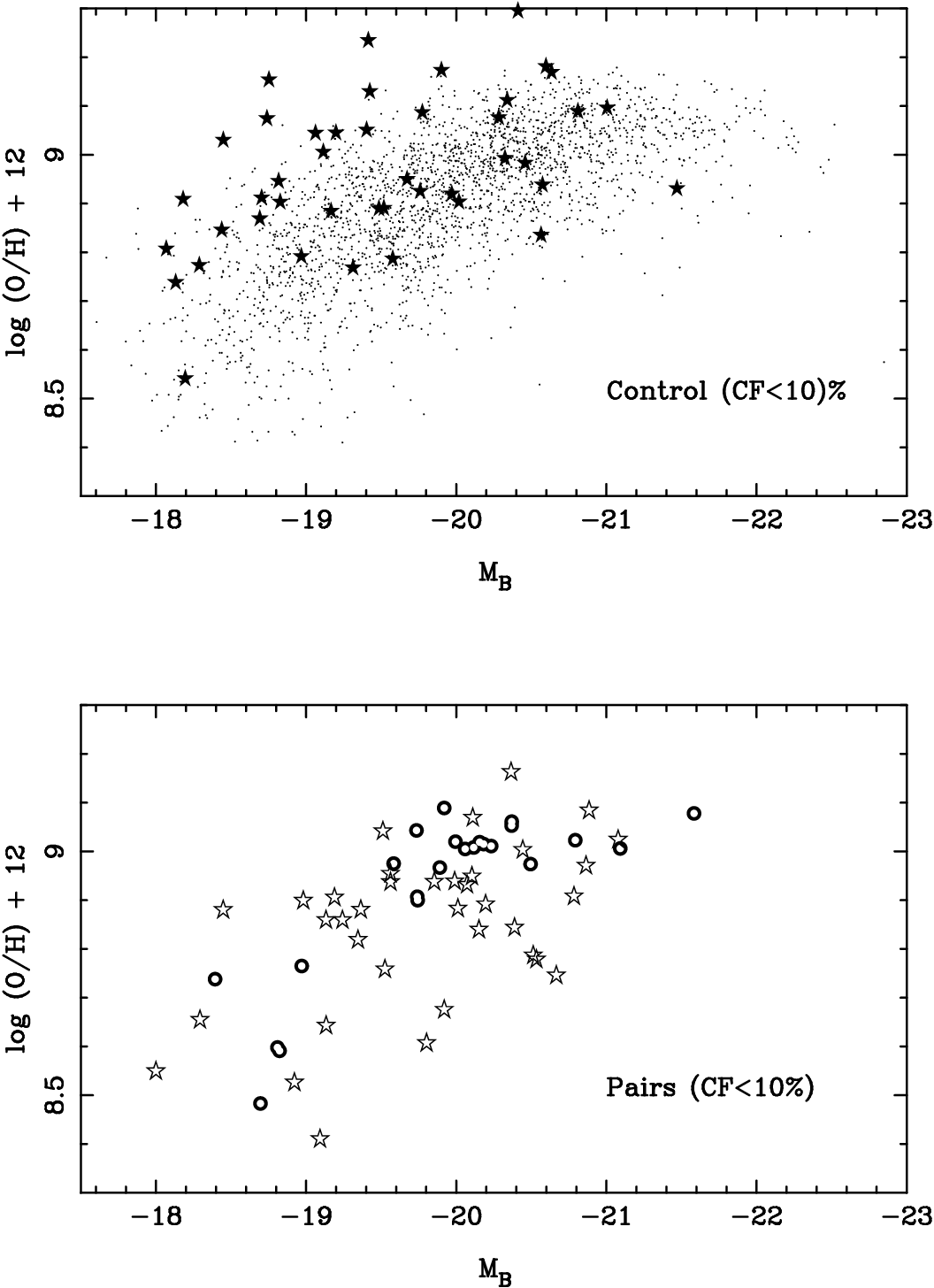}}}}
\caption{\label{compare_kewley} Luminosity-metallicity (LZ) 
relations for the control samples of Kewley et al. (2006)
and this paper (top panel) and the pairs from the same works
(bottom panel).  Symbols are as before -- filled points are
control, open points are pairs; stars are for Kewley et al. (2006)
circles/dots for the SDSS.  Only the CF$<$10\% galaxies from the SDSS
have been plotted in order to be comparable with Kewley et al.}
\end{figure*}

We next consider
whether there any obvious differences between the selection
of Kewley et al. (2006a) and our samples.  Both works rely on
pair identification from transverse (projected) separation and 
relative velocity.  We have selected our close pairs sample $r_p < 30$
\hkpc\ to match the closest separation bin of Kewley et al. (2006a).
Our velocity cut is somewhat more stringent than Kewley et al.,
500 km/s rather than 1000 km/s.  However, repeating the LZ and MZ
analyses with a 1000 km/s cut for the SDSS pairs does not change 
our results (increasing the velocity range only increases our pairs
sample by 7\%).  The CfA pairs sample has a lower redshift range
than the SDSS, the former having a lower redshift cut-off of
$z=0.0077$ and a median redshift of $z=0.018$,
which is close to the low $z$ cut-off in the SDSS.  However, we consider it
unlikely that evolutionary effects can be significant over the
redshift ranges covered by the two surveys.  Ellison \& Kewley
(2005) and Kewley \& Ellison (2008) have also stressed the importance
of using the same metallicity diagnostics in comparisons,
since there can be a factor of three offset for different calibrations.
Both Kewley et al. (2006a) and our work both use the Kewley \& Dopita (2002)
`recommended' metallicity calibration, so there should be no offset
due to diagnostic differences.  At this point, it is
instructive to compare the two control samples of this work
and Kewley et al. (2006a).  Although the selection of the CfA 
sample is done in the
$B$-band, as opposed to the $r$-band selection of SDSS pairs,
Figure \ref{compare_kewley} shows that a similar range in M$_B$
is probed by both samples (although the latter extends to slightly
more extreme values at both ends of the M$_B$ distribution, thanks to
the larger sample).  Figure \ref{compare_kewley} also shows that,
despite our caveat in \S \ref{sample_sec} that we may be missing low
metallicity galaxies, the SDSS sample is not deficient in sub-solar 
abundance galaxies compared with the CfA.
Nonetheless, from Figure \ref{compare_kewley}
it is clear that the NFGS control galaxies are inconsistent
with the SDSS control; a 2D KS test rules out the null
hypothesis with 99.8\% confidence.  Therefore, despite apparently
similar selection in terms of redshift, projected separation,
$\Delta$v, metallicity diagnostic and CF, the LZ distributions of 
the NFGS and SDSS \textit{control} samples are significantly different.

\begin{figure*}
\centerline{\rotatebox{0}{\resizebox{13cm}{!}
{\includegraphics{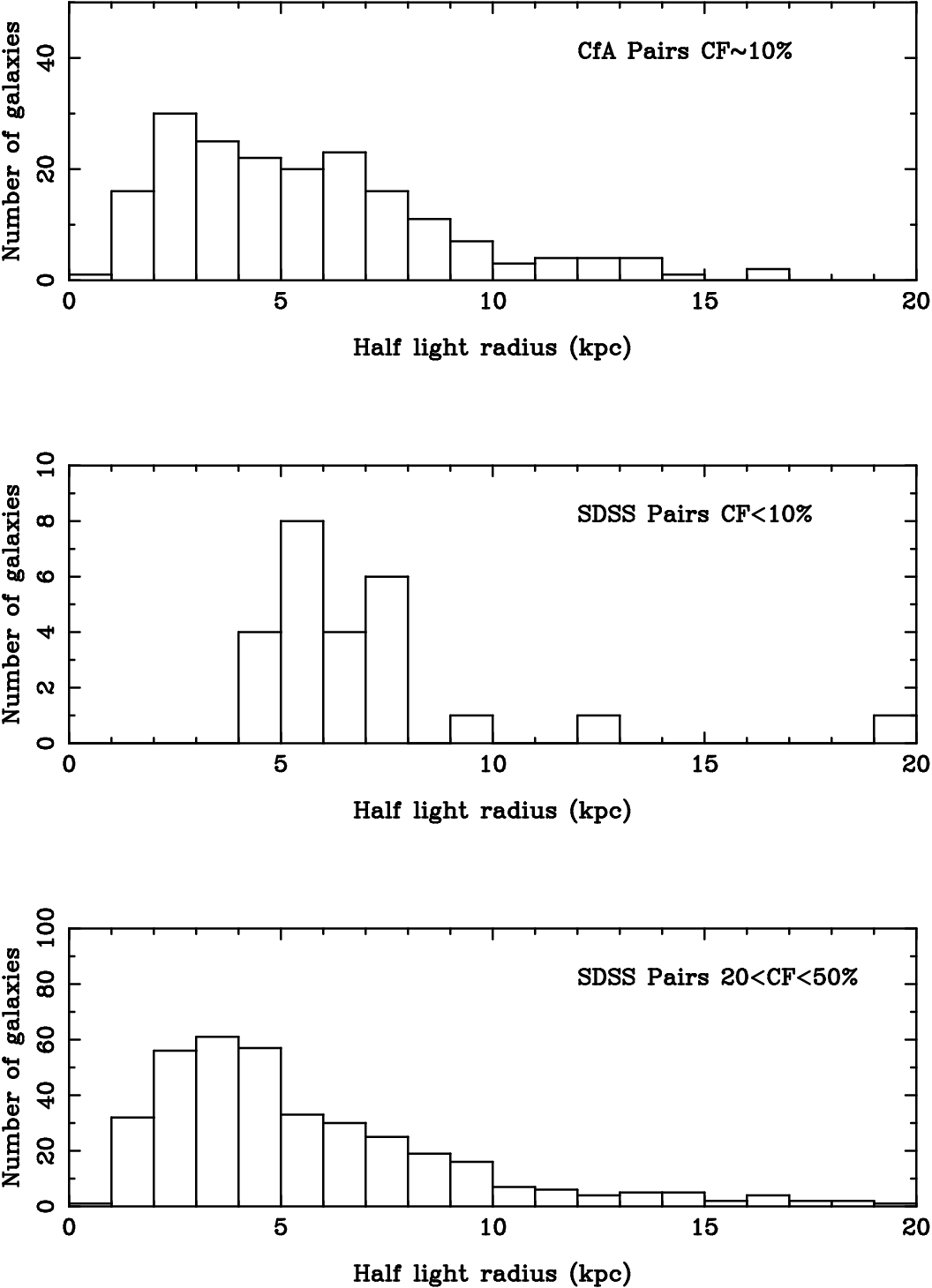}}}}
\caption{\label{rh_histo} Half light radii for the CfA pairs sample
(top panel), SDSS pairs with CF$<$10\% (middle panel) and SDSS
pairs with $20<$ CF $<50$\% (bottom panel). }
\end{figure*}

\begin{figure*}
\centerline{\rotatebox{270}{\resizebox{13cm}{!}
{\includegraphics{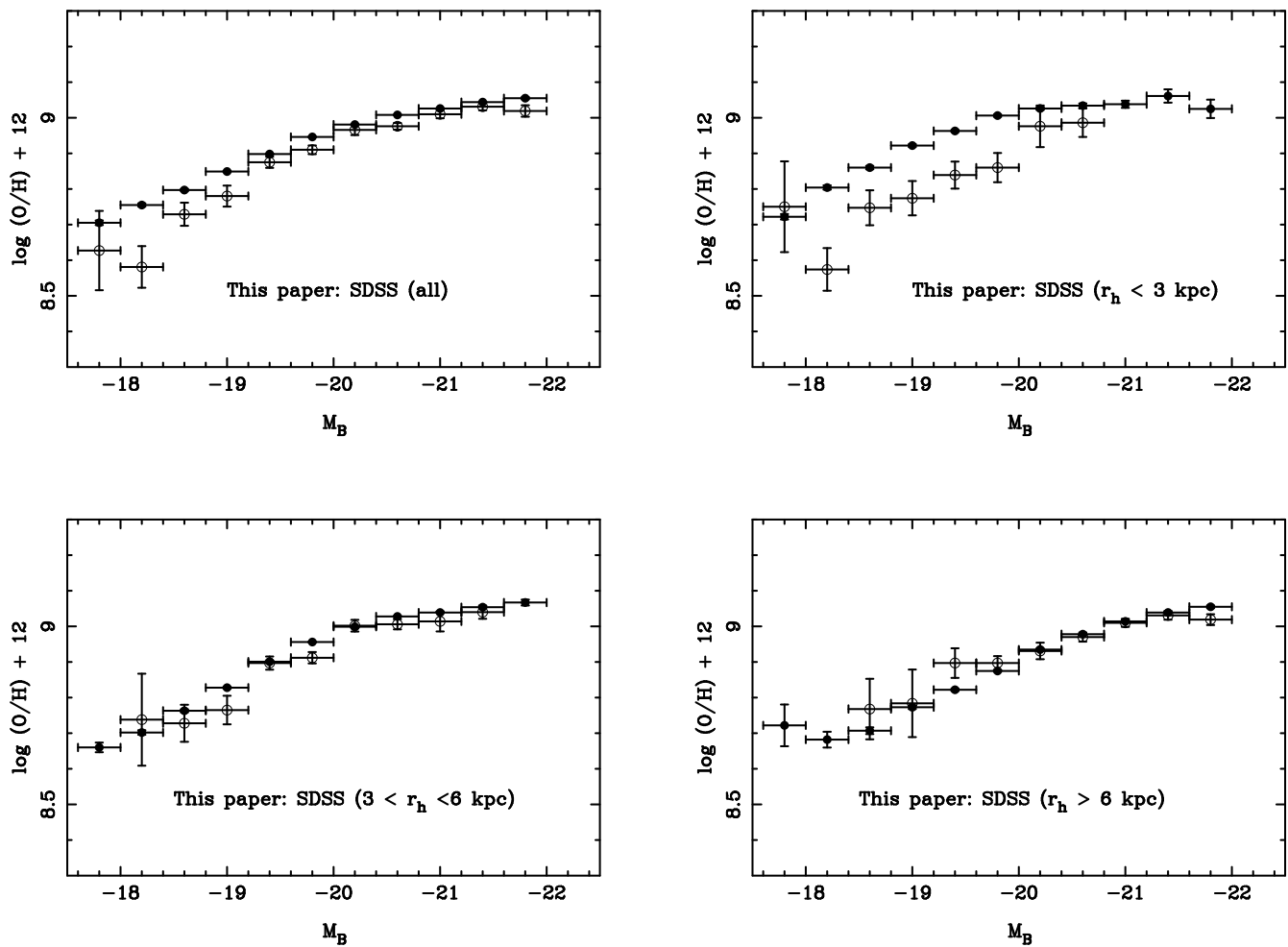}}}}
\caption{\label{LZ_rh} Luminosity-metallicity (LZ) relations for our
SDSS pairs sample. In all panels, filled points refer to 
control samples and open points to pairs. The panels show different
cuts in galaxy half light radius.  }
\end{figure*}

A possible clue as to the origin of the difference between the
CfA/NFGS and SDSS samples is revealed by the trend in LZ offset with
CF seen in Figure \ref{LZ_bin}.   Although the SDSS pairs show mildly enhanced
metallicities for CF$<$10\%, for $10<$ CF $<20$\% there is no offset
compared with the control, but at $20<$ CF $<50$\% the pairs are
systematically more metal-poor.  Since CF will obviously be
a strong function of galaxy half light radius, 
so the trend in LZ offset with CF might actually be a
trend in galaxy size.  If confirmed, this would imply that galaxies
with smaller $r_h$ tend to have low metallicities for their
luminosity/mass, whereas larger galaxies may be offset in the opposite
direction.  In Figure \ref{rh_histo} we compare the $r_h$
distributions of the CfA pairs with the SDSS pairs with two CF cuts:
CF$<$10\% and $20<$ CF $<50$\%.  The histogram clearly shows that the
CfA pairs have a $r_h$ distribution that is skewed towards smaller
sizes than the SDSS CF$<$10\% pairs.  Therefore, although these two
samples have similar covering fractions, the size distribution of
galaxies is very different.  On the other hand, the SDSS $20<$ CF
$<50$\% CF and CfA pairs have very similar $r_h$ distributions.  In
turn, the LZ relations of these two samples (CfA pairs and SDSS pairs
with $20<$ CF $<50$\%) show concordantly low metallicities for a given
luminosity.  We can see this explicitly in Figure \ref{LZ_rh} where we
plot the LZ relation for different half light radii; galaxies with
$r_h < 3$ \hkpc\ are metal-deficient for their luminosity, but this
effect is absent for larger galaxies.  The small enhancement in
metallicity that was present for small CFs in Figure \ref{LZ_bin} is
absent for the large $r_h$ sub-sample in Figure \ref{LZ_rh}.  This may
be due to the fact that nuclear spectra are required to see the
effect, i.e. the galaxies need to be large \textit{and} the spectra
must have small covering fractions.  Our sample is not large enough to
test this hypothesis, but it would be clearly interesting to obtain
more nuclear spectra of galaxies with $r_h >$ 6 \hkpc\ in the future.
Finally, the results shown in Figures \ref {LZ_bin} and \ref{LZ_rh}
also demonstrate that the impact of low metallicity gas infall is seen
not only in the CF$\sim$ 10\% nuclear spectra of the CfA pairs, but
also in the larger covering fractions of the SDSS pairs.  This
indicates that the offset in the LZ relation may be driven by changes
that occur on scales larger than `nuclear'.  There are (at least) two
reasons why this might be the case.  First, we may be observing the
galaxies early enough in their interaction that the gas flows are
still on-going, i.e.  the metal poor gas is still on its way to the
center.  This would imply that the offset in the LZ plane on scales of
several kpc is highly transient. Alternatively, galaxy interactions,
which are thought to enhance bar formation (e.g. Gerin, Combes \&
Athanassoula 1990) and contribute to central gas flows (e.g. Friedli
\& Benz 1993), may result in galaxies with flatter abundance gradients
(e.g. Martin \& Roy 1994).  Combined with the transport of metal-poor
gas to the center, this could result in a longer lasting suppression
of the LZ relation in some galaxy pairs.  We return to the reason for
the offset in the LZ relation in section \ref{shift_sec}.

\subsection{The Mass Metallicity Relation}

\begin{figure*}
\centerline{\rotatebox{270}{\resizebox{13cm}{!}
{\includegraphics{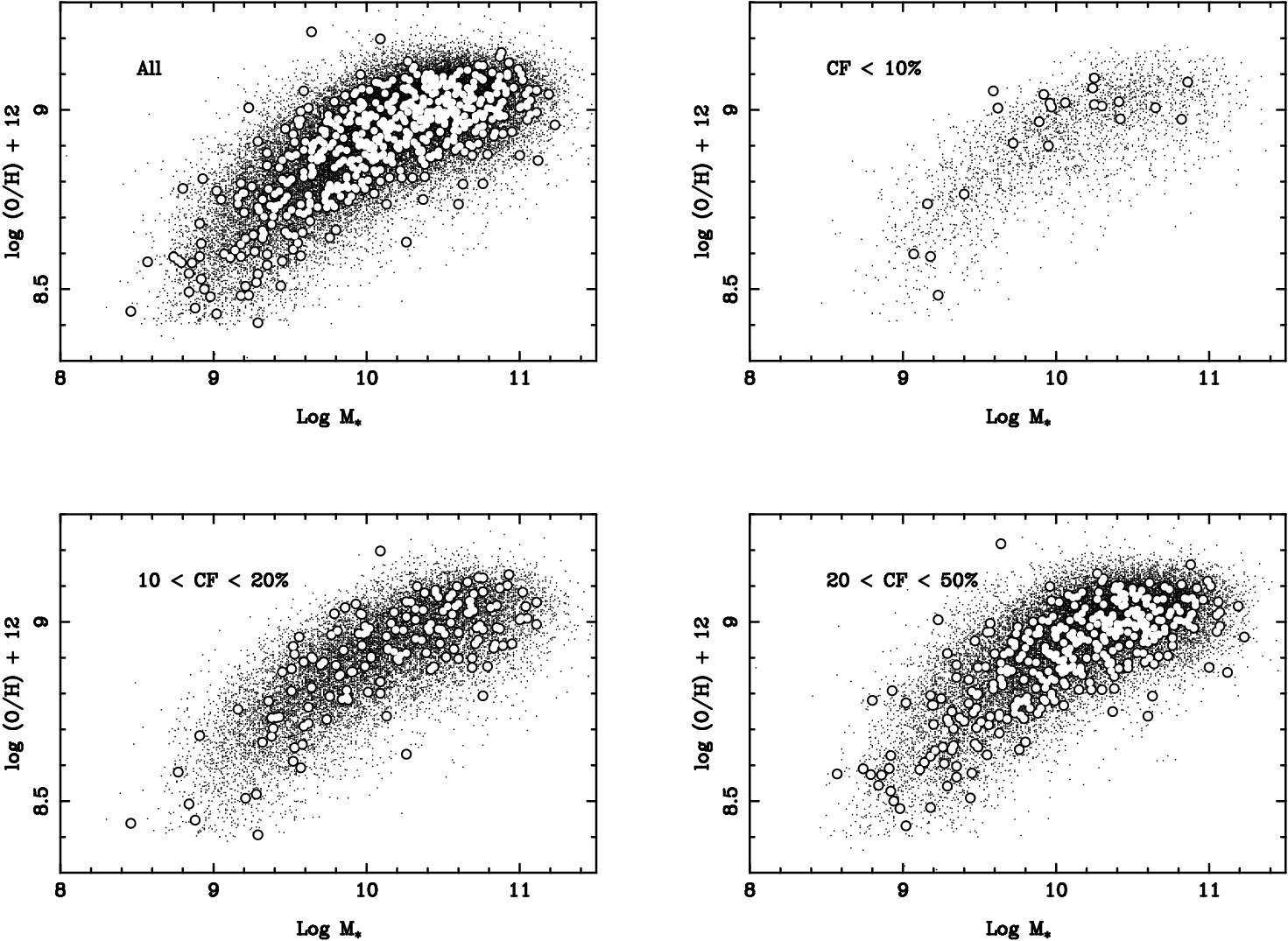}}}}
\caption{\label{MZ} Mass-metallicity (MZ) relations for our control
sample (black dots) compared with galaxies with close companions 
(open points) for various
cuts in $g$-band fiber covering fraction (CF). }
\end{figure*}

\begin{figure*}
\centerline{\rotatebox{270}{\resizebox{13cm}{!}
{\includegraphics{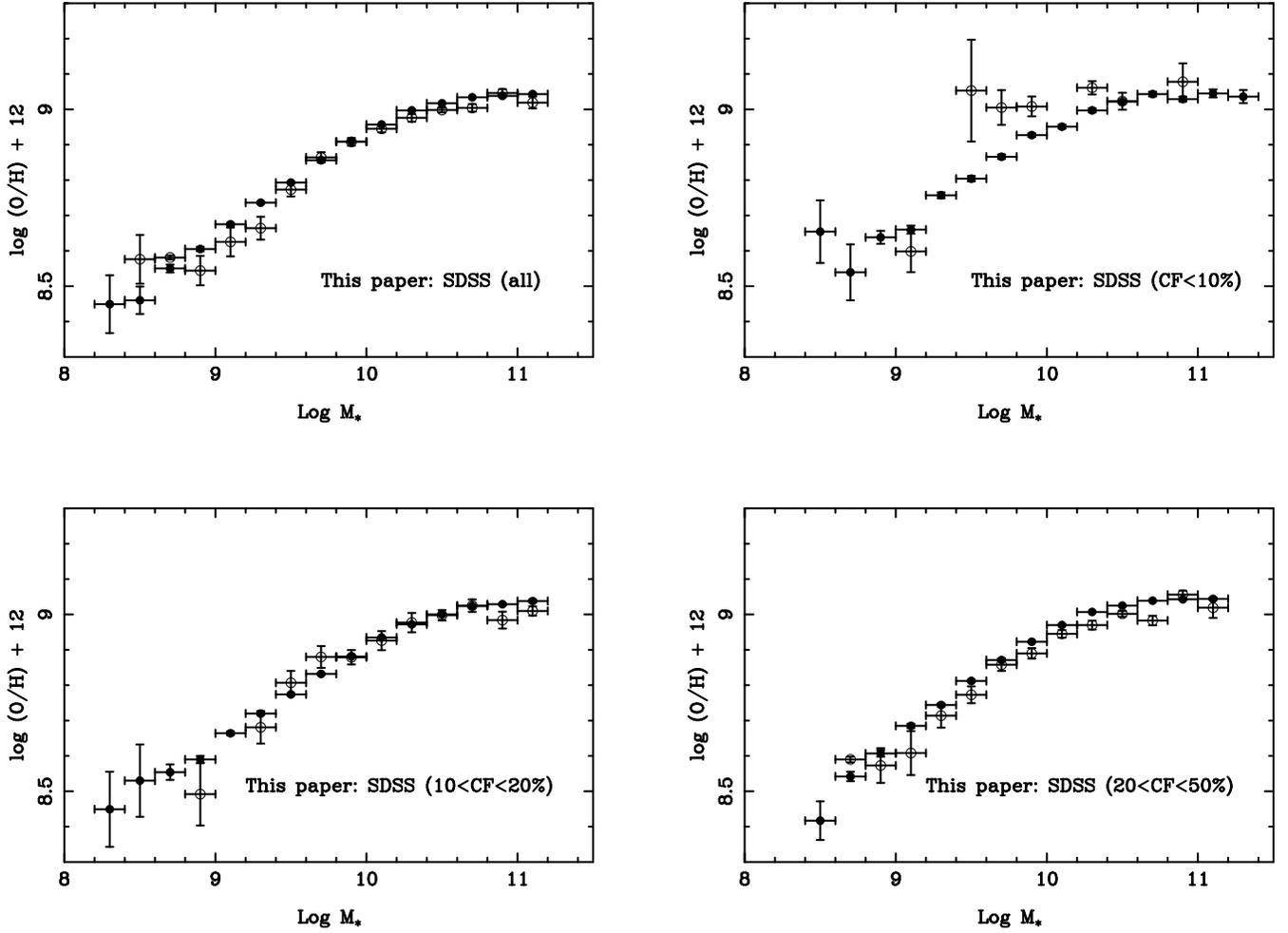}}}}
\caption{\label{MZ_bin} Mass-metallicity (MZ) relations for our control
sample (filled circles) compared with galaxies with close companions 
(open circles) for all values of CF (top panel), and CF$<$10\%
(bottom panel) binned by log stellar mass. }
\end{figure*}

We repeat the analysis of the previous section, but now
replace luminosity with stellar mass.  In Figure \ref{MZ} we show the MZ 
relation for our control and
close pairs samples for different cuts in covering fraction.  Comparison
with Figure \ref{LZ} highlights the result of Tremonti et al. (2004)
that the MZ relation is much tighter than the LZ relation, with a 
1$\sigma$ spread $<$ 0.2 dex for a given stellar mass.     
In Figure \ref{MZ_bin} we show the
binned MZ relation for close pairs and control samples 
for all the SDSS galaxies as well as for the
three CF cuts. Although there is a slight
tendency towards marginally lower metallicities for a given
mass in the full pairs sample, as seen in the binned LZ relation,
the shift is again $<$ 0.05 dex, and not significant given
the error bars.  However, the CF$<$10\% sample again shows a significant
enhancement in metallicity at intermediate masses.  The
KS probability that the MZ distributions of the CF$<$10\% pair
and control samples being drawn from the same population
is 2\%, i.e. as significant as the LZ result for the CfA sample 
(Kewley et al. 2006a) and slightly more significant than the LZ
result for the SDSS pairs presented above.  
We see a similar trend in the offset in metallicities as a function
of covering fraction in the MZ relation as in the LZ relation --
an increase in metallicity for small CFs and lower metallicities
for pairs with high CF spectra.
However, although the offset towards lower metallicities in the
20$<$ CF $<$ 50\% CF bin is systematic in MZ, it is slightly
less statistically significant than in the LZ.  
Whereas the offset in the LZ relation
for 20$<$ CF $<$ 50\% is 0.05 -- 0.1 dex, with the largest offsets at
the lowest luminosities, the offset in MZ is consistently
around 0.05 dex.  This indicates that the brightest galaxies (M$_B <
-20$) may be exhibiting a pure metallicity shift.  This is perhaps
not surprising since a starburst of fixed luminosity will have
a fractionally small impact on the luminosity of an intrinsically
bright galaxy.  Moreover, the flat slope of the LZ relation at
bright magnitudes means that any luminosity shift will need to be
large in order to be detected.  However, if the metallicity shift
is about 0.05 dex downwards for all luminosities/masses (as
indicated in Figure \ref{MZ_bin}), there may be an additional
luminosity component to the LZ relations shift that contributes
up to $\sim$ 0.4 mag.

Kewley et al. (2006a) argued that the offset observed in the LZ relation
determined for their CfA pairs sample was driven by a difference in
metallicity rather than luminosity.  Their argument was based on the fact
that their absolute magnitudes were derived from the $r$-band where
new star formation will contribute little continuum flux.  Barton
et al. (2001) also concluded that triggered star formation will not
significantly increase the luminosity of a paired galaxy, based on
comparisons of the Tully-Fisher relation.  However,
the marginally larger offset that we find in the LZ relation, particularly
at M$_B > -20$, compared to the
MZ relation for a given CF, raises the question as to whether 
some of the shift may be
due to an increased luminosity in pairs, as well as a lower metallicity.
Although there is little continuum flux expected from a starburst in
the $r$-band, the H$\alpha$ line is present in this bandpass and
may contribute significantly.  To test whether the shift in the LZ
relation may be due to increased luminosity in close galaxy pairs,
we calculate the absolute magnitude in four SDSS filters ($u, g, r$ and $i$)
and use these magnitudes in the LZ relation.  If an increase in
luminosity from a central starburst is shifting the LZ relation of pairs
towards brighter absolute magnitudes, we expect to see this effect
more strongly in the blue filters.

\begin{figure*}
\centerline{\rotatebox{270}{\resizebox{13cm}{!}
{\includegraphics{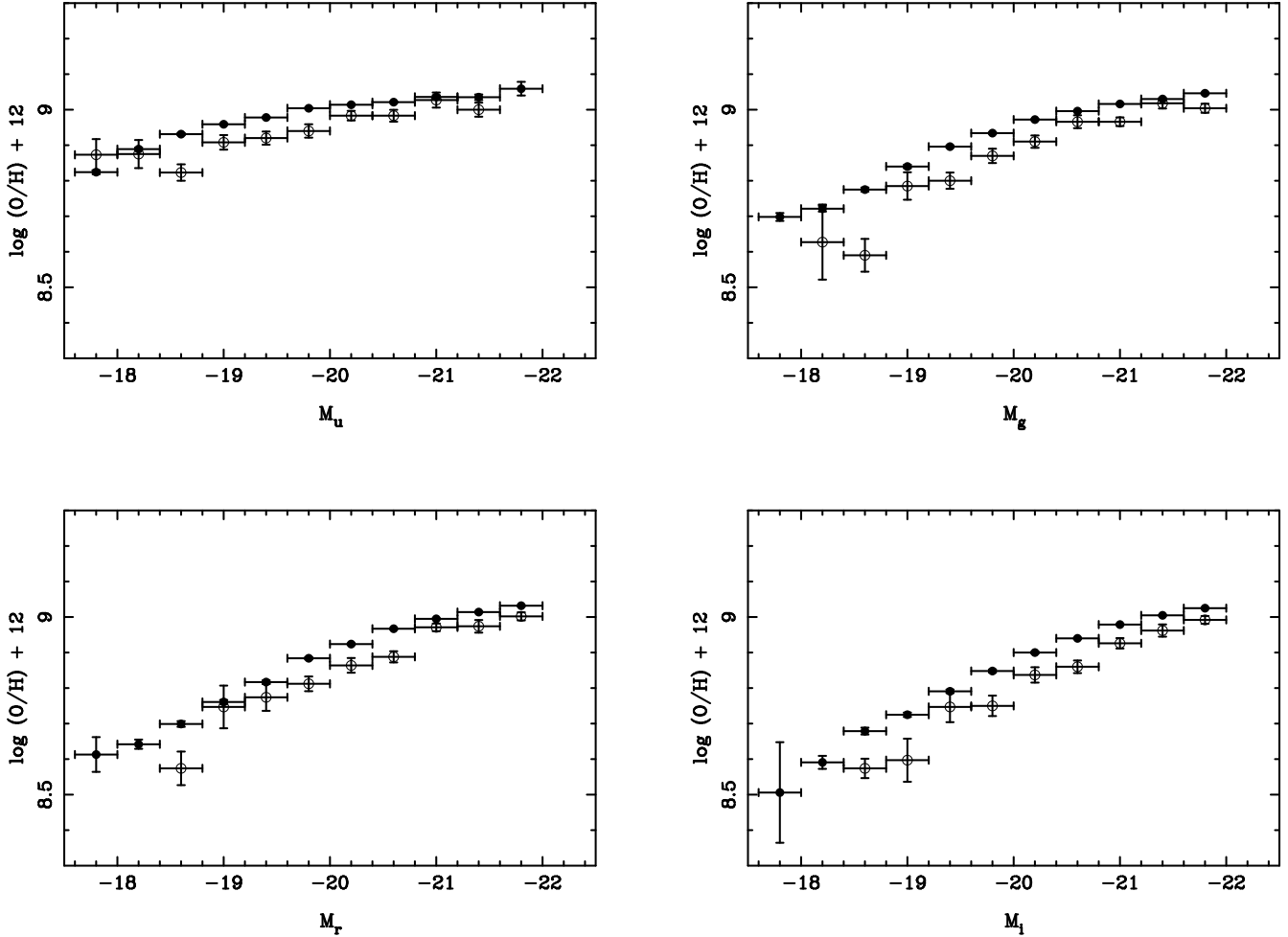}}}}
\caption{\label{LZ_ugri} Luminosity-metallicity (LZ) relations for our control
sample (filled circles) compared with galaxies with close companions 
(open circles) for $20 <$ CF $<50$\%.  The four panels show the LZ
relation as determined for absolute magnitudes in four different SDSS
filters: $u, g, r$ and $i$. The persistence of an offset in the LZ
relation even in the reddest SDSS filters indicates that the shift
is predominantly in metallicity, not higher luminositites due to increased
star formation.}
\end{figure*}

In Figure \ref{LZ_ugri} we show the LZ relation derived for 
20$<$ CF $<$ 50\% for SDSS control and pairs samples for absolute magnitudes
in four filters.  We first note that the LZ relation is much flatter
for bluer filters, an effect particularly noticeable in the $u$-band.
This is probably due to the high sensitivity of the $u$-band magnitude
to instantaneous star formation which smears out the underlying
relation of metallicity with mass.  The correlation of $i$-band magnitude
with metallicity very closely resembles the MZ relation since redder
filters more faithfully represent the underlying stellar mass.
Figure \ref{LZ_ugri} shows that the horizontal shift in the linear
(fainter absolute magnitude) part of the LZ relation is shifted 
marginally more in the $u$-band
filter ($\sim$ 0.75 mags) than in the other filters ($\sim$ 0.6 mags).
Combined with the smaller shift in the MZ relation, this  
indicates that part of the overall shift of pairs relative to
control galaxies in the LZ relation
may be due to the brightening of pairs experiencing a starburst.
This idea is further supported by the brighter median M$_B$ in the
close pairs sample: $-20.15$ compared with $-19.94$ for the
control galaxies.  Recall that the two samples are well matched in
mass (see Figure \ref{z_histo_cull}), so that this difference in
absolute magnitude is likely associated with the additional star
formation in pairs found in \S\ref{sfr_sec}.  Shioya, Bekki \&
Couch (2004) model the change in absolute magnitude in starbursting
mergers and predict a total brightening of $\sim$ 1 magnitude in
M$_B$.  However, fading happens rapidly and a brightening of
a few tenths of a magnitude is commensurate with a time of
only a few hundred million years after the burst.

\subsection{On the Shift in the LZ/MZ Relations}\label{shift_sec}

The results in the previous sections, and shown in Figures
\ref{LZ_bin} and \ref{MZ_bin}, hint that the magnitude and direction
of the offset in the LZ/MZ relations is a function of covering
fraction.  We have also shown that the dependence on CF is a
manifestation of a strong empirical dependence on the intrinsic galaxy
half light radius.  Ellison et al. (2008) have shown that a
segregation in the MZ relation exists even within the control sample
of galaxies.  However, in the control sample of non-paired galaxies
used by Ellison et al. (2008) there is a shift towards lower
metallicities for larger radii.  In the pairs sample, it is the
galaxies with the smallest half light radii that show lower
metallicities for a given mass.  The mechanism for the metallicity
shift in pairs is therefore likely to be driven by a different
physical cause.  Based on qualitatively similar downward shifts in 
metallicity for
a given stellar mass in ULIRGs and compact UVLGs (Rupke et al. 2007; Hoopes
et al. 2008), but the absence of a significant
dependence on large scale environment
(Mouhcine et al. 2007), it is likely that this effect is due to merger
activity.  In this subsection we explore two possible
`fundamental' parameters that may be the underlying cause of the $r_h$
dependence of the MZ relation for paired galaxies.

\subsubsection{Dynamical Time}

Finlator \& Dav\'e (2007) have recently proposed a general model
(i.e. not specific to galaxy pairs) for the existence and form
of the MZ relation.  These authors suggest that the MZ (and by
association, the LZ) relation 
can be understood via the interplay of gas accretion from the intergalactic
medium, star formation and subsequent mass loss through winds.
In this model, there is an equilibrium metallicity for a galaxy
of a given mass from which the galaxy may be displaced 
by the inflow of metal-poor
material.  In response to the deposition of fresh fuel, which in
turn increases the gas surface density, the galaxy will experience
an increase in its SFR.  A key parameter in this model is the
ratio of the galaxy's dynamical time ($t_{\rm dyn}$) and the dilution
time ($t_d$).  The dilution time is defined as the time taken
for the galaxy to recover from the injection of metal
poor gas, and return to its equilibrium metallicity.  If $t_d < 
t_{\rm dyn}$ then the galaxy `recovers' its equilibrium
metallicity promptly, leading to very little scatter in the MZ relation.  
Conversely, if $t_d > t_{\rm dyn}$, then the galaxy struggles to 
recover promptly from inflows.  In Figure \ref{tdyn} we test the
effect of $t_{\rm dyn}$ in the normalization of the LZ relation
by splitting the pairs and control galaxies
by dynamical time, which we calculate from $r_h$ and stellar mass.
For short dynamical times, we find a tendency for pairs to have low
metallicities for their luminosity.  This could be understood, in the
context of the model described above, if galaxies with short $t_{\rm dyn}$
are those that most efficiently funnel metal-poor gas.  However, it
is then difficult to explain why galaxies with longer dynamical times
should have metallicities higher than
the control sample.  Enhanced metallicities might be associated with
induced star formation that has already deposited its metals back
into the ISM, but this is unexpected for long $t_{\rm dyn}$, which should
have less prompt induced star formation than galaxies with short 
$t_{\rm dyn}$. We therefore conclude that dynamical time is unlikely to be the
fundamental parameter driving the sensitivity of the LZ in pairs
to $r_h$.  This is perhaps not surprising given that the gas accretion
in the `field' galaxies simulated by Finlator \& Dav\'e (2007)
occurs via a very different mode than the infall of gas to the nucleus
of a paired galaxy.  I.e. the former is dependent on the free-fall
time of gas from the intergalactic medium, whereas the latter requires
funnelling of gas to the center that is already settled in the outer part
of the galactic disk.

\subsubsection{Bulge Fraction}

The segregation of the galaxy pair LZ/MZ empirically depends on $r_h$.
Galaxies with smaller sizes for a given mass will have a higher mass
density, whereas galaxies with larger $r_h$ and the same mass will
have shallower mass potentials.  We therefore next consider whether it
is the spatial mass distribution in galaxies that drives the offset
in the LZ and MZ relations of paired galaxies.  Simulations
of galaxy interactions have previously shown that one of the
factors that regulates gas inflow and nuclear starbursts is the
relative prominence of the galaxy's bulge (e.g. Mihos \& Hernquist 
1994, 1996; Cox et al. 2007).
Bulges appear to provide stability against gas inflow, so that
galaxies with low bulge fractions more efficiently funnel gas to their
centers.  We therefore investigate whether bulge fraction
may be driving the different offsets in the LZ/MZ relations
for different galaxy half light radii.

\begin{figure*}
\centerline{\rotatebox{270}{\resizebox{13cm}{!}
{\includegraphics{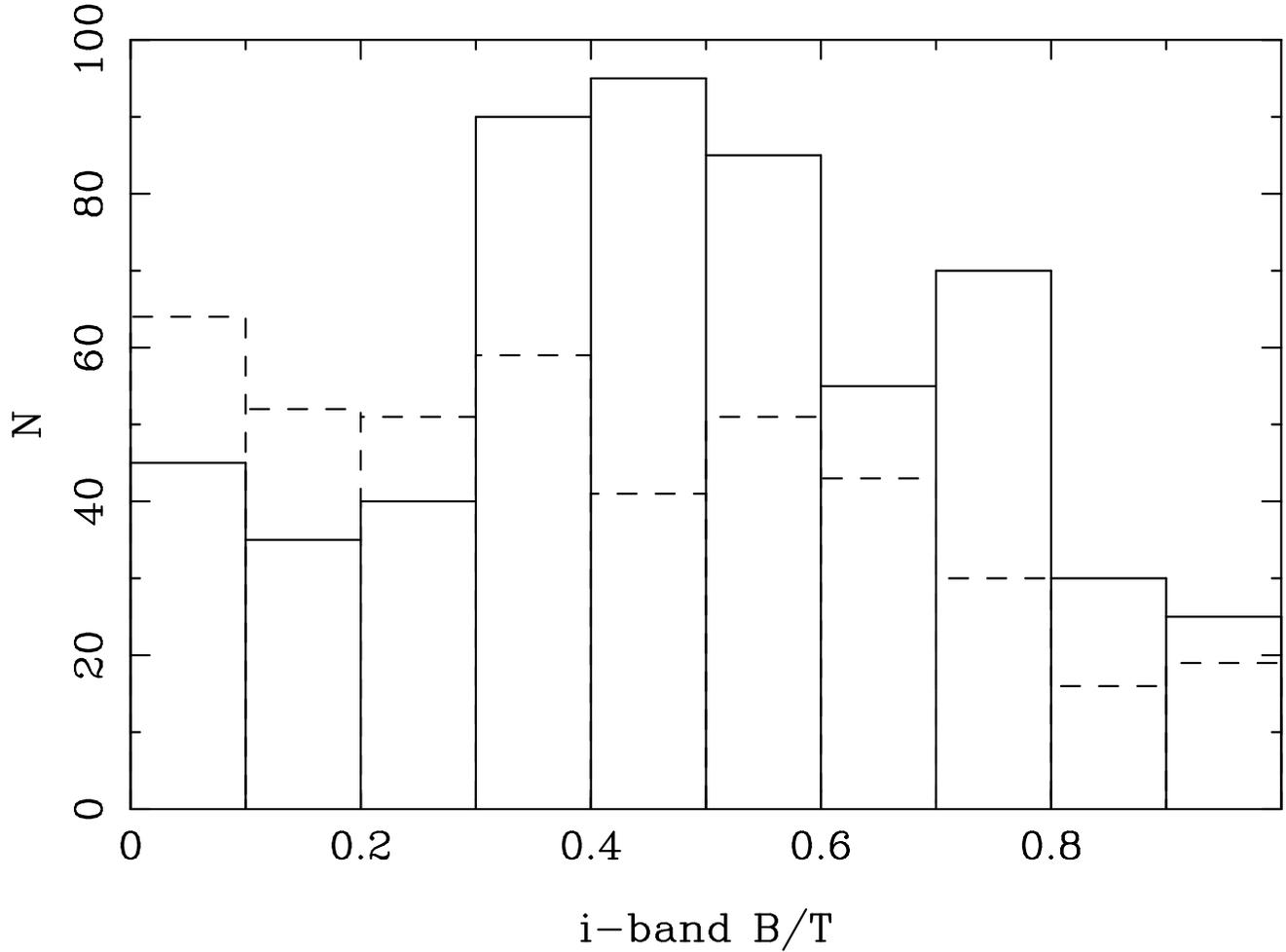}}}}
\caption{\label{bti_histo} Histogram of $i$-band B/T ratios for galaxies
in pairs.  The solid line shows galaxies with $r_h < 3$ \hkpc\ and
the dashed line is for larger galaxies.  The solid histogram has been
scaled up by a factor of 4 for display purposes.  }
\end{figure*}

In Figure \ref{bti_histo} we show the histogram of $i$-band
bulge-to-total (B/T) ratios for galaxies with close companions.  
We chose the
$i$-band for this comparison since the B/T fractions measured
in blue filters may primarily measure any increase in nuclear
star formation (e.g. Paper II).  The $i$-band is selected to be
a good indicator of the underlying mass distribution between
the bulge and the disk. In Figure \ref{bti_histo} we have
further divided the close pairs sample into those galaxies which
have small half light radii, $r_h < 3$ \hkpc, and those with
larger sizes.  Figure
\ref{bti_histo} shows that small $r_h$ galaxies in close pairs tend
to have higher bulge fractions than large galaxies.  The
KS probability that the two distributions are the same is 0.007.

\begin{figure*}
\centerline{\rotatebox{270}{\resizebox{13cm}{!}
{\includegraphics{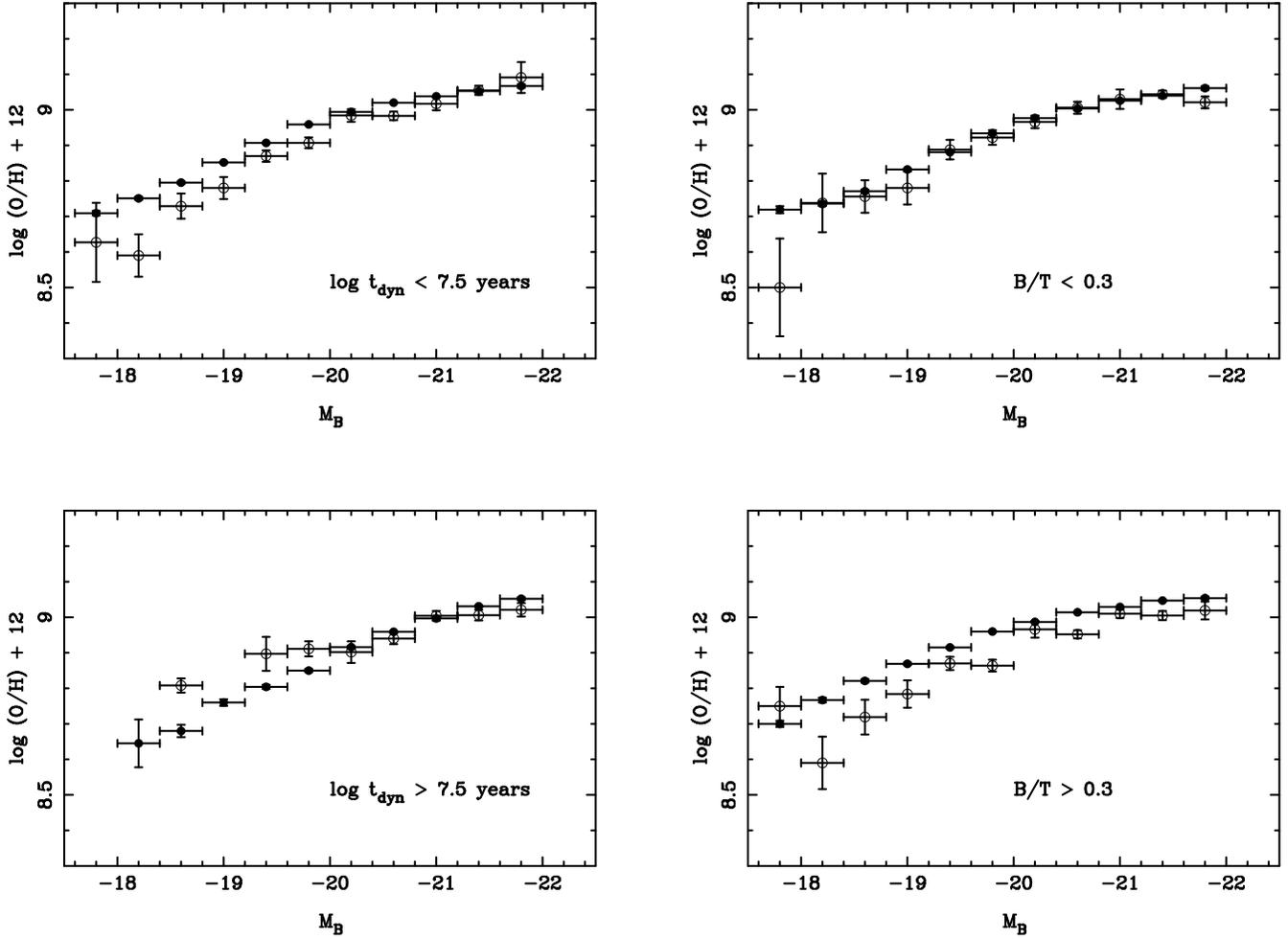}}}}
\caption{\label{tdyn} The luminosity-metallicity relation for SDSS
close pairs (open circles) and control galaxies (filled circles)
divided by dynamical time (left panels) and bulge fraction (right
panels).  }
\end{figure*}

Figure \ref{bti_histo} shows a potential link between $r_h$
and B/T.    Cox et al. (2007) have suggested that galaxies in unequal mass
mergers with B/T$>0.3$ will have 
burst efficiencies 3 times lower than a bulgeless galaxy in
an otherwise identical interaction\footnote{Cox et al. (2007)
deal with bulge-to-disk ratios, which we convert to B/T for
consistency.}.  Paired galaxies with $r_h < 3$ \hkpc\ appear
to have a marked dearth of bulge fractions below this value,
indicating that small galaxies may be less efficient at funnelling
gas to their centers for star formation. A possible explanation
for the offsets in the LZ relation seen in Figure \ref{LZ_bin}
may therefore be the connection between galaxy size and typical
bulge fraction.  Indeed, dividing the galaxy samples by B/T does
show an LZ offset for large, but not small, bulge fractions (see
Figure \ref{tdyn}).  This can be explained if smaller galaxies 
($r_h < 3$ \hkpc), which tend to have B/T $>$ 0.3 (see Figure
\ref{bti_histo}), have their
metal-poor gas reservoirs disrupted in an interaction, leading to an overall
injection of metal poor gas into the central $\sim$ 5--10 \hkpc.
However, this gas is not efficiently funnelled into the very
center of the galaxy, leading to less efficient star formation
and overall lower gas metallicity extending over a projected
area of several kpc.  Although this gas may eventually experience a starburst,
Cox et al. (2007) have shown that this event is delayed relative
to the initial (first passage) starburst by $\sim$ 1 Gyr.
Larger galaxies, which are more likely to have B/T$<0.3$,
more efficiently funnel gas to their centers, leading to
a prompt nuclear starburst and rapid metal-enrichment and recovery
to metallicity levels commensurate with the control sample
(Figure \ref{tdyn}).

\begin{figure*}
\centerline{\rotatebox{0}{\resizebox{13cm}{!}
{\includegraphics{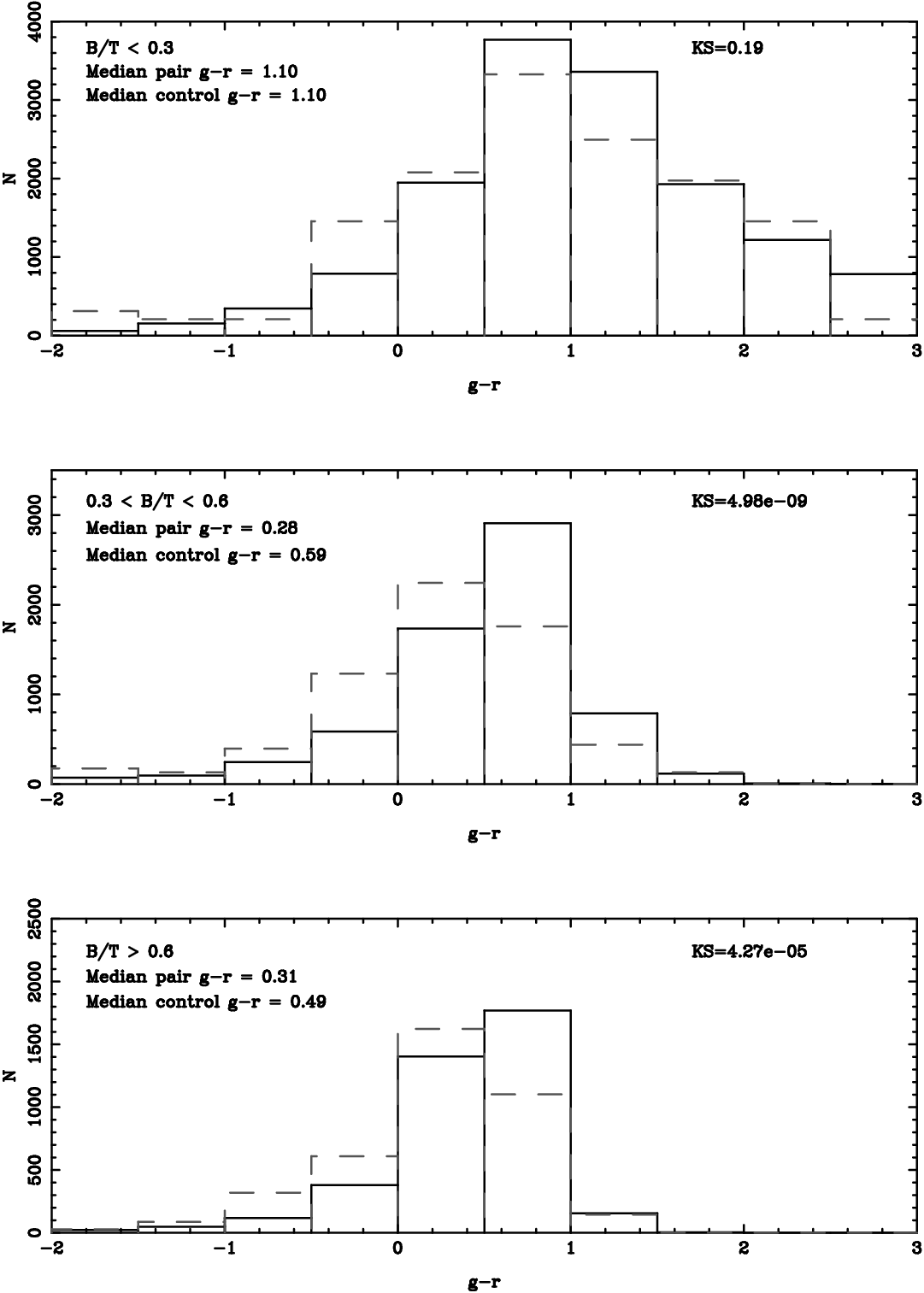}}}}
\caption{\label{gmr_bulge_bt} $g-r$ colors for control (solid histogram)
and paired (dashed histogram) galaxies for three cuts in bulge
fraction (B/T).  Each panel is labelled with the median color
of the pair and control galaxy distribution and the KS probability
that the two are drawn from the same parent population. }
\end{figure*}

To further test this hypothesis, in Figure \ref{gmr_bulge_bt}
we plot the bulge $g-r$ colors for 3 cuts in
B/T, where the cuts are applied to both the control and pairs
samples.  We find that the galaxies with the smallest
bulge fractions (B/T$<0.3$) have no difference in
$g-r$ color, compared to 0.31 and 0.18 for
$0.3<$ B/T $<0.6$ and B/T $>$ 0.6 respectively.  Indeed, the distribution
of $g-r$ colors for the lowest bulge fraction galaxies is
actually consistent (KS probability = 0.19)
between the control sample and the pairs.  The key to interpreting
this result is relative timescales: that of color changes following
a starburst versus interaction timescales.  The scenario decribed above,
in which $r_h$ depends on B/T, the latter parameter being a 
determining factor in the efficiency of nuclear star formation,
could explain the $g-r$ distributions of Figure \ref{gmr_bulge_bt}
if the timescale for post-starburst color changes is shorter than,
or comparable to, the dynamical time of the pair. 
Bruzual \& Charlot (1993) show that for a 10$^7$ year burst of
star formation, the optical colors evolve most rapidly over the
first 10$^8$ years after the starburst.  After this point, both
the models and actual star cluster data show a relative plateau
in color, changing by less than 0.1 mag in $B-V$ up to 1 gigayear.
Moreover, the fading of a starburst is typically a few magnitudes
from $10^7 - 10^8$ years after the burst, after which it will
usually be barely visible on top of the continuous, ambient star-forming
galaxy population (Sawicki, private communication), although the exact
contrast will of course depend on the relative strength of the starburst.
The typical dynamical time of close pairs is of the order of a 
few hundred Myrs to half a gigayear
(Mihos \& Hernquist 1996; Barton et al. 2000; Patton et al. 2000).
We therefore speculate that one explanation of our observations
is that many of the close pairs in our sample have already experienced
gas disruption from an initial pass $\sim 10^8$ years ago.  
In the larger galaxies (which have a tendency towards smaller bulge fractions) 
this has resulted in a prompt nuclear starburst and metal-enrichment
leading to high metallicities for a given luminosity compared with
the field, but a stellar
population that has already lost its massive O and B stars.
In the smaller galaxies, the re-distribution of metal-poor gas
has led to a lower metallicity for a given luminosity compared with
the field.  In these bulge dominated galaxies, star formation still
occurs, but is delayed relative to the first passage (Cox et al. 2007), 
so we still see the evidence of on-going activity in their colors.

\section{AGN Fraction}\label{agn_sec}

There is strong observational and theoretical evidence linking the
interactions of galaxies and the onset of nuclear activity.
Storchi-Bergmann et al. (2001) found a correlation between central
star formation activity and AGN in interacting galaxies, providing
a causal link between the two processes.  This observation was confirmed
by Kauffmann et al. (2003a) who found that the star formation in AGN
dominated galaxies is distributed over the central few kpc of active
galaxies.  Kauffmann et al. (2003a) also found that a
larger fraction of AGN galaxies (as opposed to non-active massive
galaxies) have experienced significant bursts of star formation in
the past few gigayears.  Alonso et al. (2007) draw a similar conclusion,
based on lower values of the break index $D_n$(4000) which indicates
more recent star formation in visibly merging galaxies with AGN activity.
In previous sections, we have presented evidence for central starburst
activity in close galaxy pairs; do we see any evidence for enhanced
AGN activity in our pairs that has followed the starburst?

To investigate this question, we remove the criterion that galaxies must be
classified as HII (star-forming) galaxies and also include those
that have been classified as dominated by an AGN ionizing spectrum.
The classification of galaxies as star-forming or AGN dominated
can be achieved with a variety of line strength diagnostics; in this
work we use the diagnostic of Kewley et al. (2001).  This leads
to an approximate 10\% increase in the size of our pairs and control 
samples.  However, we  impose the criterion that the
bulge-to-total ratio be in the range
$0<$ B/T $<$1, i.e. that the galaxy is fitted with two components
and excludes pure disks and pure bulges, 
which significantly reduces the number of
galaxies considered.  Furthermore, although 
our main pairs sample is still defined as containing galaxies with
companions whose separations lie in the range $r<$ 30 \hkpc, we
also consider a wide pairs sample of galaxies whose companions
have separations $30 < r < 80$ \hkpc.  The wide pairs sample acts
as a consistency check, since any differences due to proximity
should be weaker in the wide pairs sample than the sample of close
pairs.  A summary of the numbers
of galaxies in the various samples considered in this section
are given in Table \ref{AGN_frac}.

We now examine the fraction of galaxies in the pairs versus control sample
which are classified as AGN as a function of color, B/T and smoothness;
our results are given in Table \ref{AGN_frac}.
The smoothness parameter, $S$, is derived from the GIM2D bulge+disk
fits as described in detail by Simard et al. (2002).  In brief,
$S$ measures both the smoothness of the disk+bulge and its
asymmetry with higher values of $S$ indicating a higher degree
of asymmetry across the galaxy within 2 half light radii.  Smoothness is
therefore a good indictor of morphology with later type galaxies
exhibiting generally higher values of $S$ (McIntosh, Rix \& Caldwell
2004).  Here, we use $S_g$, smoothness as measured in the $g$-band.

\begin{figure*}
\centerline{\rotatebox{270}{\resizebox{13cm}{!}
{\includegraphics{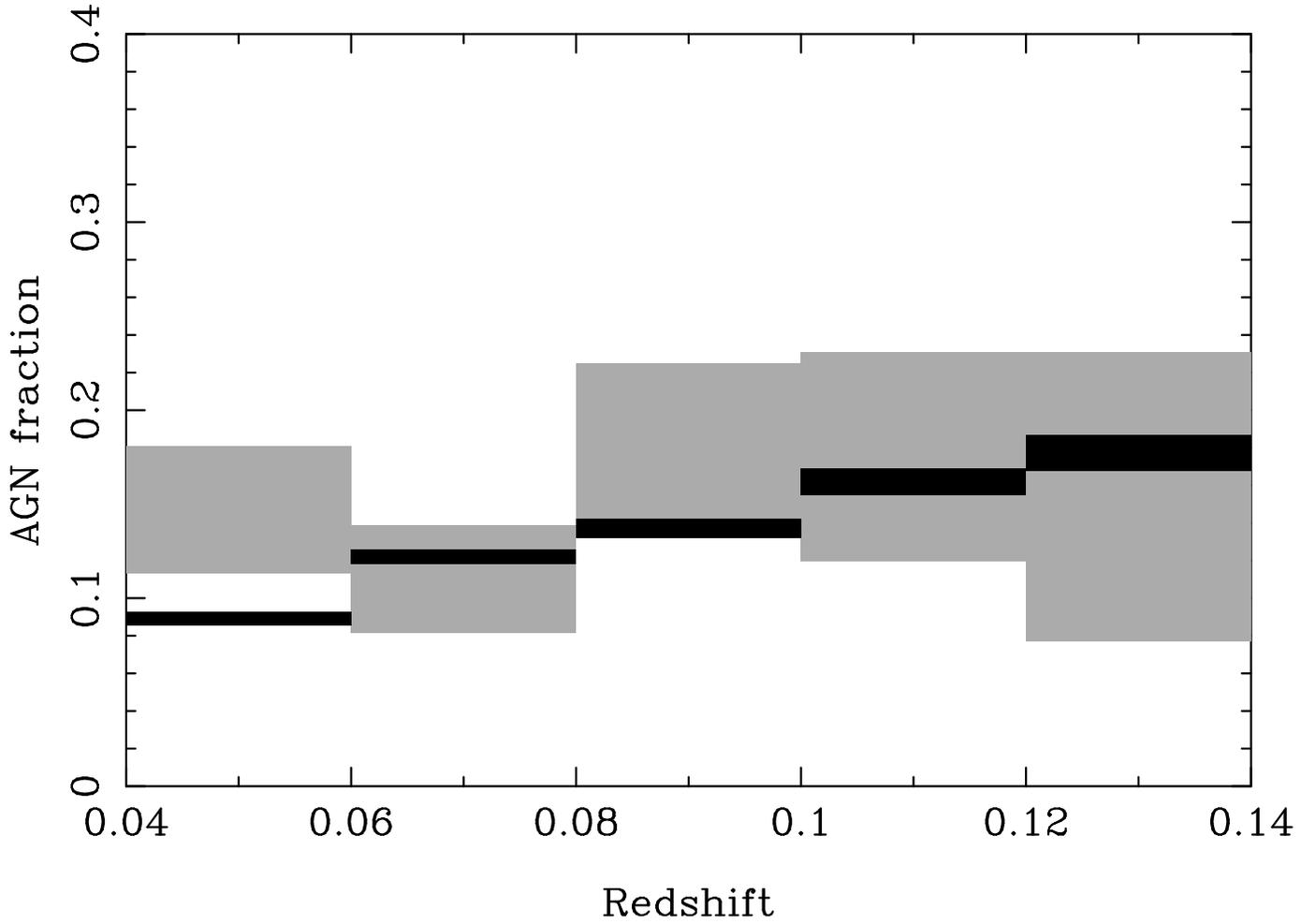}}}}
\caption{\label{agn_frac_fig} AGN fraction for all control galaxies
(filled black rectangles) and close pairs (gray shaded rectangles)
as a function of redshift.  The width of each rectangle shows the
size of the redshift bin, and the height indicates the median
AGN fraction in that bin and its spread based on Poisson statistics. }
\end{figure*}

In Figure \ref{agn_frac_fig} we show the fraction of `all' galaxies
(i.e. corresponding to the first line in Table \ref{AGN_frac}) that
are AGN as a function of redshift.  The control galaxies show a steady
increase in AGN fraction with redshift. However, this is likely to be
dominated by systematic selection rather than physical effects. Since the
stellar mass distribution is strongly skewed to higher values at
higher redshifts (see discussion in \S\ref{sample_sec}) and higher
mass galaxies have higher AGN fractions (lines 2 and 3 in Table
\ref{AGN_frac}) it is not surprising that the control galaxies exhibit
increasing AGN fraction at higher redshifts.  Restricting our sample
to only galaxies with stellar masses above 10.5 M$_{\odot}$ reverses
the trend and gives lower AGN fractions at higher redshift.  This is
likely to be due to aperture bias (e.g. Kauffmann et al. 2003a).
Despite these systematic effects, we can still compare
\textit{differentially} the AGN fraction in the control and pairs
samples as a function of redshift.  Apart from the lowest redshift
bin in Figure \ref{agn_frac_fig}, the AGN fractions are consistent
between the control and the pairs samples.  However, the small
number of pairs (particularly at high redshift) in each
redshift bin means that the uncertainties on AGN fraction are
quite high.

The results in Table \ref{AGN_frac} show that different selection
criteria yield different AGN fractions.  In general, more massive,
redder, elliptical (low $S_g$, high B/T)
 galaxies have a higher AGN fraction than
less massive, bluer, spiral galaxies.  There are a few selection
criteria for which the close pairs have a higher AGN fraction
than the control, e.g. $(g-r)_{\rm bulge} \ge 0.8$.  However,
in no case do we see a higher AGN fraction for close pairs
than for \textit{both} the wide pairs and the control samples.
The wide pairs add as a consistency check because a) they
do not show proximity induced effects such as enhanced SFR
or offset in LZ and b) we know that the wide pairs sample
is likely to be quite highly contaminated (e.g. Perez et al. 2006a).
The fractions given in Table \ref{AGN_frac} therefore do not
provide any convincing evidence that interactions lead to an
increased AGN fraction in close pairs.
A similar conclusion was reached by Barton et al. (2000)
for their CfA redshift pairs sample. A larger and more recent
study by Alonso et al. (2007) draws the same conclusion - the
distributions of properties such as color, concentration (analogous
to our B/T ratio) or morphology (measured here by $S_g$)
are indistinguishable for close pairs and control galaxy samples.
These results are consistent with the finding of Li et al. (2006)
that only one AGN in 100  has an extra neighbour within 70 kpc
compared with a control sample of non-AGN matched in mass and
redshift.
Extending this work beyond galaxy pairs, Miller et al. (2003), 
and references therein, found that AGN fraction is 
also independent of environment in groups and clusters. 
Most recently, Li et al. (2008b) have used a sample of 90,000 AGN
from the SDSS DR4 to demonstrate that although active galaxies
with close neighbours show similar enhancements in star formation
as non-AGN galaxies, the presence of close neighbours does not
promote nuclear activity.
Conversely, Woods \& Geller (2007) find a higher AGN
fraction in both minor and major pairs compared with field
galaxies in a sample of 1200 galaxies with companions in the SDSS
DR5.  If we had not considered separately the close and wide
pairs, and considered the latter as a `secondary control',
we would have drawn an identical conclusion for some of the
subsets considered in Table \ref{AGN_frac}.  However, our expectation
that the wide pairs should approximate to the control sample,
lead us to reject the significance of the increased AGN fraction
in the three cases where it is seen in Table \ref{AGN_frac}.

The typical dynamical and burst timescales of close pairs are 
typically a few hundred Myrs
(Mihos \& Hernquist 1996; Barton et al. 2000), an order of magnitude
shorter than the time-since-burst of the Kauffmann et al.
(2003a) AGN sample.  Taken together, this paints a picture of
delayed AGN activity that begins much later than the initial
central starburst.  This is also the scenario provided by
merger models which show that starbursts in the central
regions of galaxies can be seen early in the interaction
process.  However, accretion rates only increase later when the
merging is much more advanced, i.e. after at least a Gyr, and the galaxy has
formed a massive elliptical (e.g. Bekki \& Noguchi 1994;  Di Matteo,
Springel \& Hernquist 2995;  Bekki et al. 
2006).  The simulation results are born out observationally by the work
of Alonso et al. (2007) who, having found no distinction in galaxy properties
for their pairs/control sample, visually classified the subset of
pairs that were clearly interacting or merging.  This visual classification
led to a clear distinction in the properties of galaxies that were
actively merging, rather than those that were simply close in
$\Delta v$ and separation.   

\section{Summary and Conclusions}\label{summary_sec}

We have presented a sample of 1716 galaxies with close ($r_p < 80$ \hkpc,
$\Delta v < 500$ \kms\ and 0.1$<$ M$_1$/M$_2 < 10$)
companions selected from the SDSS DR4, whose properties we have
compared with a control sample of 40095 galaxies.  
The combination of photometric and spectroscopic data for these
galaxies yields a consistent, large sample of properties
including metallicity, SFR, mass, B/T ratios, colors
 and AGN contribution.  Our main conclusions are

\begin{itemize}

\item \textbf{Star Formation Rate and Proximity:} Galaxy pairs have
higher SFRs by up to 70\% for separations $<$ 40 \hkpc\ compared with
a control sample of galaxies with equal stellar mass distribution.  This
result is in agreement with inferences from numerous other studies
(e.g. Barton et al. 2000; Lambas et al. 2003;
Nikolic et al. 2004; Geller et al. 2006) which have measured the enhancement
of H$\alpha$ equivalent width as a function of separation.

\item \textbf{Star Formation Rate and Relative Galactic Stellar Mass:}  
The enhancement in SFR is largest for galaxies in pairs
with mass ratios 0.5$<$ M$_1$/M$_2 < 2$ and steadily decreases 
for paired galaxies with more discrepant stellar masses. 
We find tentative evidence 
for enhanced SFR in the less massive galaxy of a minor (mass ratio
greater than 2:1) pair, but the result is not statistically
significant.  The (luminosity-selected) pairs study of 
Woods \& Geller (2007) provides the strongest evidence for
a more enhanced SFR in the lower mass galaxy of a pair. 

\item \textbf{Luminosity- and Mass-Metallicity Relation:} We find an
offset in the LZ and MZ relations for galaxies in pairs with $r_p <
30$ \hkpc\ relative to our control sample.  For galaxies with small
half light radii ($r_h < 3$ \hkpc), which tend to be observed with
large covering fractions in the SDSS, we find a 0.05--0.1 dex offset
in the LZ relation towards lower metallicity in the pairs compared
with the control.  This is consistent with the previous result of
Kewley et al. (2006a).  A shift is also present in the MZ relation for
large CF/small $r_h$ galaxies, at the 0.05 dex level.  Based on the LZ
relation derived for absolute magnitudes in different SDSS filters we
conclude that the shift is partly in metallicity ($\sim$ 0.05 dex) and
partly in luminosity (up to 0.4 mags at M$_B >-20$).  We find
tentative evidence that larger $r_h$ galaxies (which tend to be
observed with small covering fractions) may have enhanced metallicity
for a given mass/luminosity in pairs relative to the field.  We
investigate what fundamental parameters may drive the empirical
dependence of the LZ/MZ offsets on $r_h$.  We conclude that a
dependence on bulge fraction provides a consistent picture with the
observations.  In this scenario, the smaller galaxies (with half light
radii typically $r_h <$ 3 \hkpc), tend to have larger bulges which
delays the interaction-induced star formation.

\item  \textbf{AGN Fraction:}
For given cuts in color, bulge fraction and smoothness, pairs of galaxies
have AGN fractions consistent with the field, consistent with the
conclusions of Barton et al. (2000) and Alonso et al. (2007).  
However, redder galaxies and those with more symmetric morphologies have
higher AGN fractions ($\sim$ 20--30\%) than blue or asymmetric galaxies
($\sim$5--10\%). 

\end{itemize}

Overall, our results support the picture that close interactions
(within a few tens of kpc)
between galaxies causes gas to inflow to the central regions,
engaging new star formation.  The outer parts of the galaxy and
the disk, are largely unaffected by additional star formation 
(e.g. Paper II).  
The process of gas infall and star formation is most efficient
for approximately equal mass galaxies, and in interactions of
galaxies with low bulge fractions.  The observed shift in the
LZ/MZ relations depends on the relative timescales of the
interactions, gas flows and induced star formation.  Based on
the consistency of AGN fraction in the pairs and control samples,
we conclude that all of these processes occur on timescales
shorter than the activation of the central black hole.

\acknowledgements

We are indebted to both the SDSS team for their provision of quality,
public datasets, and also to the Munich group for making public the
results of their spectral template fitting.  This work could not have
been done without the hard work and community spirit of the many
people who provided these catalogs.  We are particularly grateful to
Jarle Brinchmann who provided advice and guidance on using the
Munich catalogs.  We are also grateful to Lisa Kewley for providing
the metallicities used in this paper and to Betsy Barton for providing
data on the CfA pairs sample and for several useful discussions.
We also benefitted from discussions with TJ Cox, Marcin Sawicki,
Evan Skillman and Christy Tremonti.
SLE, LS and DRP acknowledge the receipt of NSERC Discovery Grants
which funded this research.

Funding for the SDSS and SDSS-II has been provided 
by the Alfred P. Sloan Foundation, the Participating Institutions, 
the National Science Foundation, the U.S. Department of Energy, the 
National Aeronautics and Space Administration, the Japanese 
Monbukagakusho, the Max Planck Society, and the Higher Education 
Funding Council for England. The SDSS Web Site is http://www.sdss.org/.

The SDSS is managed by the Astrophysical Research Consortium for the 
Participating Institutions. The Participating Institutions are the 
American Museum of Natural History, Astrophysical Institute Potsdam, 
University of Basel, University of Cambridge, Case Western Reserve 
University, University of Chicago, Drexel University, Fermilab, the 
Institute for Advanced Study, the Japan Participation Group, 
Johns Hopkins University, the Joint Institute for Nuclear Astrophysics, 
the Kavli Institute for Particle Astrophysics and Cosmology, the 
Korean Scientist Group, the Chinese Academy of Sciences (LAMOST), 
Los Alamos National Laboratory, the Max-Planck-Institute for Astronomy 
(MPIA), the Max-Planck-Institute for Astrophysics (MPA), New Mexico 
State University, Ohio State University, University of Pittsburgh, 
University of Portsmouth, Princeton University, the United States 
Naval Observatory, and the University of Washington.

\clearpage

% TABLES

%\begin{landscape}
\begin{deluxetable}{lcccccc}
\tablewidth{8.0in}
\rotate
\tablecaption{\label{AGN_frac}AGN Fractions} 
\tablehead{
\colhead{Selection criteria} &
\multicolumn{3}{c}{Number of Galaxies} & %[.2ex] \cline{2-4}
\multicolumn{3}{c}{AGN Fraction} \\ [.2ex]\cline{2-4} \cline{5-7}
\colhead{} &
\colhead{Control} &
\colhead{Wide Pairs} &
\colhead{Close Pairs} &
\colhead{Control} &
\colhead{Wide Pairs} &
\colhead{Close Pairs} \\
}
\startdata
 & &  \\
All galaxies & 30052 & 936 & 502 &  0.12$\pm$ 0.01 &  0.14$\pm$ 0.02 & 0.13$\pm$ 0.02 \\ 
Log M$_{\star} < 10.2$ M$_{\odot}$ & 13059 & 432 & 237 &  0.02$\pm$ 0.01 &  0.03$\pm$ 0.01 &  0.00$\pm$ 0.01 \\
Log M$_{\star} \ge 10.2$ M$_{\odot}$ & 16993 & 504 & 265 &  0.20$\pm$ 0.01 &  0.24$\pm$ 0.02 &  0.25$\pm$ 0.03 \\ 
$(g-r)_{\rm bulge} < 0.8$ & 14644 & 543 & 337 &  0.08$\pm$ 0.01 &  0.09$\pm$ 0.01 &  0.08$\pm$ 0.02 \\ 
$(g-r)_{\rm bulge} \ge 0.8$ & 15408 & 393 & 165 &  0.16$\pm$ 0.01 &  0.22$\pm$ 0.02 &  0.24$\pm$ 0.04 \\ 
$(g-r)_{\rm disk} < 0.5$ & 14173 & 441 & 257 &  0.04$\pm$ 0.01 &  0.07$\pm$ 0.02 & 0.07$\pm$ 0.02 \\
$(g-r)_{\rm disk} \ge 0.5$ & 15879 & 495 & 245 &  0.19$\pm$ 0.01 &  0.21$\pm$ 0.02 & 0.20$\pm$ 0.03 \\ 
B/T $<$0.2 & 14041 & 388 & 124 &  0.05$\pm$ 0.01 &  0.06$\pm$ 0.02 & 0.07$\pm$ 0.03 \\ 
B/T$\ge$0.2 & 16011 & 548 & 378 &  0.18$\pm$ 0.01 &  0.20$\pm$ 0.02 & 0.15$\pm$ 0.02 \\ 
$S_g < 0.1$ & 18856 & 518 & 213 &  0.15$\pm$ 0.01 &  0.20$\pm$ 0.02 & 0.19$\pm$ 0.03 \\ 
$S_g \ge 0.1$ & 11196 & 418 & 289 &  0.08$\pm$ 0.01 &  0.07$\pm$ 0.01 & 0.09$\pm$ 0.02 \\
$S_g < 0.1$, $(g-r)_{\rm bulge} \ge$ 0.8  & 9614 & 220 & 69 &  0.19$\pm$ 0.01 &  0.29$\pm$ 0.04 &  0.33$\pm$ 0.07 \\
$S_g < 0.1$, $(g-r)_{\rm disk} \ge 0.5$,  & 4584 & 105 & 29 &  0.33$\pm$ 0.01 &  0.47$\pm$ 0.07 &  0.52$\pm$ 0.14 \\
$(g-r)_{\rm bulge} \ge 0.8$ & & & & & & \\
\enddata
\end{deluxetable}
%\end{landscape}

% FIGURES

\clearpage

\clearpage

\clearpage

\clearpage

\clearpage

\clearpage

\clearpage

\clearpage

\clearpage

\clearpage

\clearpage

\clearpage

\clearpage

\clearpage

\clearpage

\clearpage

\clearpage


\begin{thebibliography}{}


\bibitem[Alonso et al. (2006)]{alo06}
        Alonso, M. S.,  Lambas, D. G., Tissera, P. B., Coldwell, G.,
	 2006, MNRAS, 367, 1029

\bibitem[Alonso et al. (2007)]{alo07}
        Alonso, M. S.,  Lambas, D. G., Tissera, P. B., Coldwell, G.,
	  2007, MNRAS, 375, 1017

\bibitem[Alonso et al. (2004)]{alo04}
        Alonso, M. S., Tissera, P. B., Coldwell, G., Lambas, D. G.,
	 2004, MNRAS, 352, 1081

\bibitem[Balogh et al. 1999]{bal99}
        Balogh, M., Morris, S., Yee, H. K. C., Carlberg, R. G.,
	Ellingson, E., 1999, ApJ, 527,54

\bibitem[Balogh et al. 1998]{bal98}
        Balogh, M., Schade, D., Morris, S., Yee, H. K. C., Carlberg, R. G.,
	Ellingson, E., 1998, ApJ, 504, 75

\bibitem[Barnes \& Hernquist (1992)]{bh92}
        Barnes, J. E., \& Hernquist L., 1992, ARA\&A, 30, 705

\bibitem[Barnes \& Hernquist (1996)]{bh96}
        Barnes, J. E., \& Hernquist, L., 1996, ApJ, 471, 115

\bibitem[Barton et al. (2008)]{bar08}
        Barton, E. J., Arnold, J. A., Zentner, A. R., Bullock, J. S., 
	Wechsler, R. H., 2008, ApJ, accepted

\bibitem[Barton et al. 2001]{bar01}
        Barton, E. J., Geller, M. J., Bromley, B. C., van Zee, L., 
	Kenyon, S. J., 2001, AJ, 121, 625

\bibitem[Barton, Geller \& Kenyon (2000)]{bgk00}
        Barton, E. J., Geller, M. J., \& Kenyon, S. J., 2000, ApJ, 530, 660

\bibitem[Bekki \& Noguchi (1994)]{bn94}
        Bekki, K., Noguchi, M.,  1994, A\&A, 290, 7

\bibitem[Bekki, Shioya \& Whiting (2006)]{bsw06}
        Bekki, K., Shioya, Y., \& Whiting, M., 2006, MNRAS, 371, 805

\bibitem[Blanton \& Berlind 2007]{bbh07}
        Blanton, M. R., Berlind. A. A., 
	2007, ApJ, 664, 791

\bibitem[Bresolin (2004)]{bre07}
         Bresolin, F., 2007, ApJ, 656, 186

\bibitem[Bresolin, Garnett \& Kennicutt (2004)]{bgk04}
         Bresolin, F., Garnett, D. R., \& Kennicutt, R. C.,
	 2004, ApJ, 615, 228

\bibitem[Brinchmann et al. 2004]{bri04}
        Brinchmann, J., Charlot, S., White, S. D. M., Tremonti, C., 
	Kauffmann, G., Heckman, T., Brinkmann, J.,2004, MNRAS, 351, 1151  

\bibitem[Brinchmann \& Ellis (2000)]{be02} 
        Brinchmann, J., \& Ellis, R. S.,  2000, ApJ, 536, L77

\bibitem[Brooks et al. (2007)]{bro07}
        Brooks, A. M., Governato, F., Booth, C. M., Willman, B., 
	Gardner, J. P., Wadsley, J., Stinson, G., Quinn, T.,
	 2007, ApJ, 655, L17

\bibitem[Bruzual \& Charlot (1993)]{bc93}
        Bruzual, A. G., \& Charlot, S., 1993, ApJ, 405, 538

\bibitem[Bushouse (1987)]{bus87}
        Bushouse, H. A., 1987, ApJ, 320, 49

\bibitem[Byrd \& Valtonen (1990)]{bv90}
	Byrd, G., Valtonen, M., 1990, ApJ, 350, 89

%\bibitem[Cardelli, Clayton \& Mathis (1989)]{ccm89}
%        Cardelli, J. A., Clayton, G. C., \& Mathis, J. S., 1989,
%        ApJ, 345, 245


\bibitem[Carlberg et al. (1994)]{carl94}
        Carlberg, R. G., Pritchet, C. J., Infante, L., 1994, ApJ, 435, 540

\bibitem[Cox et al. (2007)]{cox07}
        Cox, T. J., Jonsson, P., Somerville, R. S., Primack, J. R.,
	Dekel, A., MNRAS, 2007, submitted, arXiv:0709.3511

\bibitem[de la Rosa et al. (2007)]{dlr07}
        de la Rosa, I. G., de Carvalho, R. R., Vazdekis, A., Barbuy, 
	B., 2007,  AJ, 133, 330

\bibitem[Denicolo, Terlevich \& Terlevich (2002)]{dtt02}
        Denicolo, G., Terlevich, R., \& Terlevich, E.,
	2002, MNRAS, 330, 69

\bibitem[Di Matteo et al. (2005)]{dim05}
        Di Matteo, T., Springel, V., Hernquist, L., 2005, Nature, 433, 604

\bibitem[Diaferio et al. 2001]{dia01}
        Diaferio, A., Kauffmann, G., Balogh, M. L., White, S. D. M., 
	Schade, D., Ellingson, E., 2001, MNRAS, 323, 999

\bibitem[Dressler 1980]{dres80}
        Dressler, A., 1980, ApJ, 236, 351

\bibitem[Drory et al. (2004)]{dro04}
        Drory, N., Bender, R., Hopp, U.,  2004, ApJ, 616, L103

\bibitem[Ellison \& Kewley (2005)]{ek05}
        Ellison, S. L., Kewley, L. J.,	2005, pg 53,
	Proceedings of "The Fabulous Destiny of Galaxies; 
	Bridging the Past and Present", Eds Le Brun, Mazure, 
	Arnouts, Burgarella

\bibitem[Ellison et al. (2008)]{sle08}
         Ellison, S. L., Patton, D. R., Simard, L., McConnachie, A. W.,
	 2008, ApJ, 672, L107

\bibitem[Ferrarese \& Ford (2005)]{ff05}
        Ferrarese, L., \& Ford, H., 2005, Space Science Reviews, 116, 523

\bibitem[Finlator \& Dave (2007)]{fd07}
        Finlator, K., \& Dav\'e, R., 2007, MNRAS, submitted, arXiv:0704.3100

\bibitem[Friedli \& Benz (1993)]{fb93}
        Friedli, D., \& Benz, W., 1993, A\&A, 268, 65

\bibitem[Geller et al. (2006)]{gel06}
        Geller, M. J., Kenyon, S. J., Barton, E. J., Jarrett, T. H., 
	Kewley, L. J., 2006, AJ, 132, 2243

\bibitem[Gerin, M.; Combes, F.; Athanassoula, E. 1990]{gca90}
        Gerin, M., Combes, F., Athanassoula, E.,
	 1990, A\&A, 230, 37

\bibitem[Gomez et al. (2003)]{gom03}
        Gomez, P., et al., 2003, ApJ, 584, 210

%\bibitem[Hopkins et al. (2001)]{hop01}
%        Hopkins, A. M., Connolly, A. J., Haarsma, D. B., Cram, L. E.,
%	 2001, AJ, 122, 288

\bibitem[Hoopes et al. (2008)]{hoo08}
        Hoopes, C. G., et al. 2008, ApJS, in press, astro-ph/0609415

\bibitem[Hummel (1981)]{hum81}
        Hummel, E., 1981, A\&A, 96, 111

\bibitem[Kauffmann et al. (2003)]{kau03a} %agn
	Kauffmann, G., et al., 2003a, MNRAS, 346, 1055

\bibitem[Kauffmann et al. (2003)]{kau03b} %sdss masses
	Kauffmann, G., et al., 2003b, MNRAS, 341, 33

\bibitem[Kennicutt et al. (1987)]{ken87}
        Kennicutt, R. C., Jr., Roettiger, K. A., Keel, W. C., van der Hulst, 
	J. M., Hummel, E., 1987, AJ, 93, 1011

\bibitem[Kewley \& Dopita (2002)]{kd02}
        Kewley, L. J., \& Dopita, M. A., 2002, ApJS, 142, 35

\bibitem[Kewley \& Ellison (2008)]{ke08}
        Kewley, L. J., \& Ellison, S. L., 2008, ApJ, accepted

\bibitem[Kewley et al. (2006a)]{kewl06a}
         Kewley, L. J., Geller, M. J., Barton, E. J.,
	  2006a, AJ, 131, 2004

\bibitem[Kewley et al. (2006b)]{kewl06b}
         Kewley, L. J., Groves, B., Kauffmann, G., \& Heckman, T.,
	  2006b, MNRAS, 372, 961

\bibitem[Kewley et al (2001)]{kew01}
        Kewley, L. J., Heisler, C. A., Dopita, M. A., Lumsden, S.,
	2001, ApJS, 132, 37

\bibitem[Kewley et al. (2005)]{kjg05}
         Kewley, L. J., Jansen, R. A., Geller, M. J., 
	  2005, PASP, 117, 227

\bibitem[Kobulnicky, Kennicutt \& Pizagno (1999)]{kkp99}
        Kobulnicky, H. A., Kennicutt, R. C., Jr., Pizagno, J. L.,
	1999, ApJ, 514, 544

\bibitem[Kobulnicky \& Kewley (2004)]{KK04} 
        Kobulnicky, H. A., \& Kewley, L. J., 2004, ApJ, 617, 240 

%\bibitem[Kobulnicky et al. (2003)]{k03}
%        Kobulnicky, H. A., Willmer, C. N. A., Weiner, B. J., Koo, D. C., 
%	Phillips, A. C., Faber, S. M., Sarajedini, V., Simard, L.,
%	Vogt, N., 2003, ApJ, 599, 1006

\bibitem[Koppen, Weidner \& Kroupa (2007)]{kwk07}
        Koppen, J., Weidner, C., Kroupa, P., 2007, MNRAS, 375, 673

\bibitem[Lambas et al. (2003)]{lam03}
        Lambas, D. G., Tissera, P. B., Alonso, M. S., Coldwell, G.,
	2003, MNRAS, 346, 1189

\bibitem[Larson \& Tinsley (1978)]{lt78}
        Larson, R. B., \& Tinsley, B. M.,  1978, ApJ, 219, 46

\bibitem[Lee et al. (2006)]{lee06}
        Lee, H., Skillman, E. D., Cannon, J. M., Jackson, D. C., 
	Gehrz, R. D., Polomski, E. F., Woodward, C. E.,
	 2006, ApJ, 647, 970

\bibitem[Lewis et al (2002)]{lew02}
        Lewis, I., et al. 2002, MNRAS, 334, 673

\bibitem[Li et al. (2008a)]{li08a}
        Li, C., Kauffmann, G., Heckman, T. M., Jing, Y. P., White, S. D. M.,
	2008a, MNRAS submitted, arXiv:0712.3792v1

\bibitem[Li et al. (2008b)]{li08b}
        Li, C., Kauffmann, G., Heckman, T. M., White, S. D. M.,
	Jing, Y. P., 2008b, MNRAS submitted, arXiv:0712.0383v1

\bibitem[Li et al. (2006)]{li06}
        Li, C., Kauffmann, G., Wang, L., White, S. D. M., Heckman, T. M., 
	Jing, Y. P., 2006, MNRAS, 373, 457 

\bibitem[Maier et al. 2005]{mai05}
        Maier, C., Lilly, S. J., Carollo, C. M., Stockton, A., Brodwin, M.,
	2005, ApJ, 634, 849

\bibitem[Marconi et al. (2004)]{mar04}
        Marconi, A., Risaliti, G., Gilli, R., Hunt, L. K., Maiolino, R., 
	Salvati, M., 2004, MNRAS, 351, 169

\bibitem[Martin \& Roy 1994]{mr94}
        Martin, P., \& Roy, J.-R.,  1994, ApJ, 424, 599

\bibitem[Masjedi et al. (2006)]{mas06}
        Masjedi, M., et al.,  2006, ApJ, 644, 54

\bibitem[Mastropietro et al (2005)]{mas05}
        Mastropietro, C., Moore, B., Mayer, L., Wadsley, J., Stadel, J.,
	2005, MNRAS, 363, 509

\bibitem[McGaugh (1991)]{m91}
        McGaugh, S. S., 1991, ApJ, 380, 140

\bibitem[McIntosh, Rix \& Caldwell (2004)]{mrc04}
       McIntosh, D. H., Rix, H.-W., Caldwell, N., 2004, ApJ, 610, 161

\bibitem[Mendes de Oliveira et al. (2005)]{mdo05}
        Mendes de Oliveira, C., Coelho, P., González, J. J., Barbuy, B.,
	2005, AJ, 130, 55

\bibitem[Mihos \& Hernquist (1994)]{mh94}
        Mihos, C., \& Hernquist, L., 1994, ApJ, 425, L13

\bibitem[Mihos \& Hernquist (1996)]{mh96}
        Mihos, C., \& Hernquist, L., 1996, ApJ, 464, 641

\bibitem[Miller et al. (2003)]{mill03}
       	Miller, C. J., Nichol, R. C., Gomez, P. L., Hopkins, A. M., 
	Bernardi, M., 2003, ApJ, 597, 142

\bibitem[Moore et al. (1999)]{moo99}
        Moore, B., Lake, G., Quinn, T., Stadel, J., 1999, MNRAS, 304, 465

\bibitem[Mouhcine, Baldry \& Bamford (2007)]{mbb07}
        Mouhcine, M., Baldry, I. K., Bamford, S. P., 2007, MNRAS
	382, 801

\bibitem[Nikolic, Cullen \& Alexander (2004)]{nca04}
        Nikolic, B., Cullen, H., Alexander, P., 2004, MNRAS, 355, 874

%\bibitem[Osterbrock(1989)]{Osterbrock89}
%         Osterbrock, D. E. 1989, Astrophysics of Gaseous Nebulae and
%         Active Galactic Nuclei (Mill Valley; University Science Books)

\bibitem[Pagel et al. (1979)]{pag79}
        Pagel, B. E. J., Edmunds, M. G., Blackwell, D. E., Chun, M. S., 
	Smith, G., 1979, MNRAS, 189, 95

\bibitem[Patton \& Atfield (2008)]{pa08}
        Patton, D. R., \& Atfield, J. E. 2008, ApJ, submitted

\bibitem[Patton et al. (2000)]{pat00}
        Patton, D. R., Carlberg, R. G., Marzke, R. O., Pritchet, C. J., 
	da Costa, L. N., Pellegrini, P. S., 2000, ApJ, 536, 153

\bibitem[Patton et al. (2005)]{pat05}
         Patton, D. R., Grant, J. K., Simard, L., Pritchet, C. J., 
	 Carlberg, R. G., Borne, K. D., 2005, AJ, 130, 2043

\bibitem[Patton et al. (1997)]{pat97}
        Patton, D. R., Pritchet, C. J., Yee, H. K. C., Ellingson, E., 
	Carlberg, R. G., 1997, ApJ, 475, 29

\bibitem[Perez et al. (2006)]{per06a}  % SFR
        Perez, M. J., Tissera, P. B., Lambas, D. G., Scannapieco, C.,  
	2006a, A\&A, 449, 23

\bibitem[Perez et al. (2006)]{per06b} % metals
        Perez, M. J., Tissera, P. B., Scannapieco, C., Lambas, D. G., 
	de Rossi, M. E., 2006b, A\&A, 459, 361

\bibitem[Pettini \& Pagel (2004)]{pp04}
        Pettini, M., \& Pagel, B. E. J., 2004, MNRAS, 348, L59

\bibitem[Pimbblet et al (2002)]{pim02}
        Pimbblet, K, Smail, I., Kodama, T., Couch, W., Edge, A., 
	Zabludoff, A.,  O' Hely, E., 2002, MNRAS, 331, 333

\bibitem[Poggianti et al. 1999]{pog99}
        Poggianti, B., Smail, I., Dressler, A., Couch, W.,
	Barger, A., Butcher, H., Ellis, R.,  Oemler, A., 1999, ApJ, 518, 576

\bibitem[Proctor et al. (2004)]{proc04}
        Proctor, R. N., Forbes, D. A., Hau, G. K. T., Beasley, M. A., 
	De Silva, G. M., Contreras, R., Terlevich, A. I., 
	 2004, MNRAS, 349, 1381

\bibitem[Rupke, Veilleux \& Baker (2007)]{rvb07}
        Rupke, D. S. N., Veilleux, S., \& Baker, A. J., 2007, ApJ,
	in press, arXiv:0708.1766

\bibitem[Salzer et al. (2005)]{sal05}
         Salzer, J. J., Lee, J. C., Melbourne, J., Hinz, J. L., 
	 Alonso-Herrero, A., Jangren, A., 2005, ApJ, 624, 661

\bibitem[Schmitt (2001)]{sch01}
        Schmitt, H. R.,  2001, AJ, 122, 2243

\bibitem[Shankar et al. (2004)]{sha04}
        Shankar, F., Salucci, P., Granato, G. L., De Zotti, G., Danese, L.,
	2004, MNRAS, 354, 1020

\bibitem[Shioya, Bekii \& Couch (2004)]{sbc04}
        Shioya, Y., Bekki, K., \& Couch, W. J., 2004, ApJ, 601, 654

\bibitem[Simard et al. (2002)]{sim02}
        Simard, L., Willmer, C. N. A., Vogt, N. P., Sarajedini, V. L., 
	Phillips, A. C., Weiner, B. J., Koo, D. C., Im, M., 
	Illingworth, G. D., Faber, S. M., 2002, ApJS, 142, 1

\bibitem[Smith et al. (2007)]{smi07}
         Smith, B. J., Struck, C., Hancock, M., Appleton, P. N., 
	 Charmandaris, V., Reach, W. T., 2007, AJ, 133, 791

\bibitem[Sorrentino, Radovich \& Rifatto (2006)]{srr06}
        Sorrentino, G., Radovich, M., Rifatto, A., 2006, A\&A, 451, 809

\bibitem[Stauffer (1982)]{sta82}
        Stauffer, J. R., 1982, ApJ, 262, 66

\bibitem[Storchi-Bergmann et al. (2001)]{sto01}
        Storchi-Bergmann, T., Gonzalez-Delgado, R. M., Schmitt, H. R., 
	Cid-Fernandes, R., Heckman, T., 2001, ApJ, 559, 147	

\bibitem[Strauss et al. (2002)]{str02}
        Strauss, M. A., et al., 2002, AJ, 124, 1810

\bibitem[Tremonti et al. 2004]{2113} 
        Tremonti, C., et al., 2004, ApJ, 693, 898

\bibitem[Wake et al. (2005)]{wake05}
        Wake, D. A., Collins, C. A., Nichol, R. C., Jones, L. R., Burke, D. J.,
	2005, ApJ, 627, 186

\bibitem[Whitmore, Gilmore, \& Jones (1993)]{wgj93}
         Whitmore, B. C., Gilmore, D. M.,  \& Jones, C., 1993, ApJ, 407, 489

\bibitem[Woods \& Geller (2007)]{fwg07}
        Woods, D. F., Geller, M. J., 
	2007, AJ, 134, 527

\bibitem[Woods et al. (2006)]{fw06}
        Woods, D. F., Geller, M. J., Barton, E. J.,
	2006, AJ, 132, 197

\bibitem[Zaritsky, Kennicutt \& Huchra (1994)]{zkh94}
        Zaritsky, D., Kennicutt, R. C., Jr., Huchra, J. P., 1994,
	ApJ, 420, 87

\bibitem[Zheng et al. (2007)]{zh07}
         Zheng, Z. X., et al., 2007, ApJ, 661, L47

\end{thebibliography}
\end{document}